\documentclass{elsart3p}

\usepackage{natbib}

\usepackage{amssymb,graphicx,rotating}

\begin{document}

\begin{frontmatter}

\title{From discs to planetesimals I: evolution of gas and dust discs}
\author{Richard Alexander\thanksref{leiden}}
\ead{rda@strw.leidenuniv.nl}
\address{JILA, University of Colorado, Boulder, CO 80309-0440, USA}
\thanks[leiden]{Present address: Leiden Observatory, Universiteit Leiden, Niels Bohrweg 2, 2300 RA, Leiden, the Netherlands}

\begin{abstract}
I review the processes that shape the evolution of protoplanetary discs around young, solar-mass stars.  I first discuss observations of protoplanetary discs, and note in particular the constraints these observations place on models of disc evolution.  The processes that affect the evolution of gas discs are then discussed, with the focus in particular on viscous accretion and photoevaporation, and recent models which combine the two.  I then discuss the dynamics and growth of dust grains in discs, considering models of grain growth, the gas-grain interaction and planetesimal formation, and review recent research in this area.  Lastly, I consider the so-called ``transitional'' discs, which are thought to be observed during disc dispersal.  Recent observations and models of these systems are reviewed, and prospects for using statistical surveys to distinguish between the various proposed models are discussed.
\end{abstract}


\end{frontmatter}

In this chapter I will discuss the processes that shape the evolution of discs around young stars.  Such discs form as a consequence of angular momentum conservation during the star formation process \citep[see, e.g., the reviews by][]{shu87,mo07}, and much of the stellar mass is thought to be accreted through the surrounding disc.  Moreover, such discs are presumed to be the sites of planet formation, and as such the study of disc evolution has important consequences for theories of both star and planet formation.  For reasons of length, in this chapter I will limit myself to the discussion of discs around stars of approximately solar mass (the T Tauri stars), but I note in passing that similar physics applies to discs observed across the stellar mass spectrum.  In the context of this volume, this chapter falls between the more general discussion of disc physics by Lodato, and the discussion of planet formation and the planet-disc interaction by Klahr.  More generally, many excellent reviews of this topic can be found in the {\it Protostars \& Planets} series (especially in the two most recent volumes), and the interested reader may also wish to consult the recent review by \citet{armitage07}.

In Section \ref{sec:obs} I discuss observations of protoplanetary discs, and the constraints these observations place on models of disc evolution.  In Section \ref{sec:gas} I consider the processes which affect the evolution of the gaseous component of protoplanetary discs, and focus in particular on models of disc photoevaporation and viscous evolution.  I then go on to consider the evolution of the solid component of the disc (in Section \ref{sec:dust}), discussing the processes of grain growth and planetesimal formation, and the dynamical processes that affect dust grains.  Lastly, in Section \ref{sec:trans} I discuss so-called ``transitional'' discs, which are believed to be crucial to our understanding of disc clearing.

\section{Observational Review}\label{sec:obs}
\subsection{Observations of protoplanetary discs}\label{sec:disc_obs}
Discs around young stars were first directly observed in the mid-1980s, initially in the form of dusty debris discs such as those seen around Vega \citep{aumann84} and $\beta$ Pic \citep{st84}, and later in the form of the gas-rich discs that we now term ``protoplanetary'' \citep{sb87}.  Advances in telescope technology in the intervening two decades has resulted in the detection, both directly and indirectly, of many more discs around young stars, and these objects are now understood to be commonplace.  Here I summarize the various methods of observing protoplanetary discs, and the constraints these observations place on models of disc evolution.

Young low-mass stars are traditionally classified in two different ways: by the shape of their infrared spectral energy distribution (SED), usually measured from multi-band photometry, or by the strength of emission lines seen by (usually optical) spectroscopy.  The SED classification scheme was proposed by \citet{lada87} and updated by \citet{am94}, and classifies objects based on the slope of their SEDs at wavelengths longward of 2$\mu$m.  Light from the central (proto-)star is absorbed by dust in the circumstellar environment and subsequently re-emitted at longer wavelengths, so a ``redder'' SED, resulting from more material in the circumstellar environment, is thought to be indicative of an earlier evolutionary phase.  The scheme is now supported by much theoretical work, and while it is recognised that the orientation of the source with respect to the observer can have an effect on the observed SED \citep[e.g.,][]{whitney03}, the existence of a broad correlation between SED class and evolutionary phase is widely accepted.  Thus Class II objects are inferred to be stars with surrounding discs, while Class III sources (with near-stellar SEDs) are assumed to be more evolved objects which have shed their discs.

The second classification scheme relies on measuring the strength of emission lines such as H$\alpha$.  Young, solar-type stars with very strong emission lines were observed as long ago as the 1940s, and as the number of known such objects increased they were classified as T Tauri stars (henceforth TTs), after their ``prototype'' T Tau \citep[][see also the review by \citealt{bertout89}]{joy45}.  It was soon realised that these were young, newly-formed stars, and that the bright emission lines (and also the bright UV emission characteristic of TTs) were produced by the accretion of material on to the stellar surface.  Later, TTs were also discovered to be bright X-ray sources \citep[e.g.,][]{kc79,fk81}, but the observations which detected TTs in X-rays also detected a further population of sources.  These were bright in X-rays but did not show the strong emission lines associated with accretion.  However, they were spatially co-located with the TTs and showed similar stellar properties, and so were christened ``weak-lined'' T Tauri stars \citep[e.g.,][]{walter88}\footnote{Note that I keep to the modern convention of ``weak-lined'' T Tauri stars, rather than ``naked'' T Tauri stars as used by \citet{walter88}.}.  We now understand that the original class of objects, now referred to as ``classical'' T Tauri stars (henceforth CTTs), represent young stars which are accreting from a circumstellar disc, while the weak-lined T Tauri stars (WTTs) are similar objects which have shed their discs.  Thus Class II objects and CTTs are inferred to possess discs, while Class III objects and WTTs are disc-less.  In modern astronomical literature these terms are used almost interchangeably, but while there is a large overlap between the two classification schemes there is not a one-to-one correspondence: a few CTTs show Class III SEDs, while a few WTTs have Class II SEDs \citep[e.g.,][]{strom93,kaas04}.  (For further discussion of disc SEDs see the chapter in this volume by Wood.)

As mentioned above, some of the most commonly observed characteristics of protoplanetary discs are signatures of gas accretion.  TTs are known to have strong magnetic fields \citep[$\sim1$kG][]{basri92,jk99}, and consequently it is believed that the inner edges of TT discs are truncated by the stellar magnetic field.  Inside this magnetospheric radius (typically $\sim10$ stellar radii) material is channelled along the magnetic field lines, and essentially free-falls on to the stellar surface \citep{gl78,hhc94}.  Typical infall velocities are 100-300km s$^{-1}$, and consequently we see a shock (known as the accretion shock) where the infalling material impacts the stellar surface.  The accretion shock typically has a temperature of $\simeq10,000$K \citep{cg98,gul00}, much hotter than the stellar photosphere, and results in an ultraviolet (UV) excess that can be measured either via the veiling\footnote{The term ``veiling'' describes the the ``filling in'' of absorption lines in the stellar spectrum by excess continuum emission, in this case from the accretion shock.} of photospheric absorption lines \citep{heg95}, or by direct measurements of the UV luminosity \citep{gul98,gul00}.  Observations of the UV excess essentially measure the accretion luminosity, and if the stellar parameters (mass, radius \& temperature) are known such observations allow the measurement of the accretion rate on to the stellar surface: typical accretion rates for CTTs range from $10^{-7}$--$10^{-10}$M$_{\odot}$yr$^{-1}$ \citep{heg95,gul98}.  In addition the infalling gas produces very broad emission lines, which are easily distinguished from the much narrower emission lines produced by the stellar chromosphere.  Observations of such lines also allow the measurement of accretion rates \citep[e.g.,][]{muz00} and, while accretion rates measured in this manner are somewhat more prone to systematic uncertainties those derived from UV continuum emission (as the relationships between line fluxes and accretion rates are empirically calibrated), they allow us to detect accretion at levels lower than can currently be observed in the UV \citep[e.g.,][]{muz03,subu05}.  Observations of accretion are important as they are the easiest way of detecting gas in protoplanetary discs, but carry the important caveat that they only probe the innermost regions ($\lesssim0.1$AU) of the disc.

\begin{figure}[t]
	\begin{center}
        \resizebox{\hsize}{!}{
        \includegraphics[angle=270]{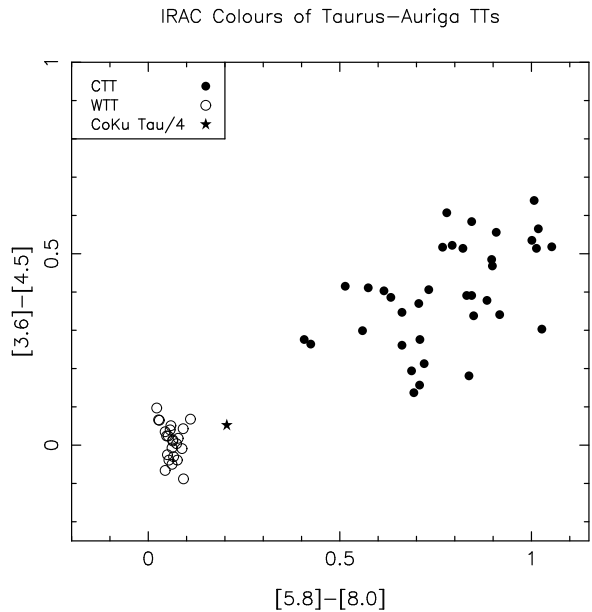}
        }
	\end{center}
        \caption{{\it Spitzer} IRAC 2-colour plot for TTs in Taurus-Auriga \citep[data from][]{hartmann05}.  The units are magnitudes: stellar photospheres have colours of zero, while red excesses have positive colours.  CTTs (classified by H$\alpha$ equivalent widths) are shown as solid circles, WTTs as open circles, and the transition object CoKu Tau/4 (see Section \ref{sec:trans}) as a star.  Note the clear gap between the loci of the Class II CTTs and Class III WTTs.}
        \label{fig:hartmann05}
\end{figure}

Perhaps the most common means of observing protoplanetary discs is the detection of infrared emission from warm dust (typically a few hundred K) in the inner part of the disc \citep[typically within $\simeq1$AU of the central star,][]{kh87,cg97}.  The resulting excess emission is most commonly observed at wavelengths from 2--10$\mu$m, and is readily detected by photometric observations \citep[e.g.,][]{kh95,haisch01,hartmann05}.  Consequently, photometric surveys of nearby star-forming regions are capable of detecting hundreds of protoplanetary discs, and large surveys of this type allow a rapid census of the disc population.  A typical example of such a survey \citep[of the nearby Taurus-Auriga star-forming region;][]{hartmann05} is shown in Fig.~\ref{fig:hartmann05}: the distinction between the disc-bearing Class II objects and the disc-less Class III sources is clear.  

Gas in discs is much harder to observe.  Warm gas in the inner disc has been observed through the detection of molecular lines, such as CO fundamental lines in the infrared \citep[][see also the review by \citealt{najita_ppv}]{najita03}, or electronic transitions of H$_2$ observed in the UV \citep{herczeg04}.  However, to date such studies have generally been restricted to relatively small numbers of objects, and as such these data are of limited use when considering disc evolution.  At larger radii, where the disc is much colder, gas is similarly hard to detect.  Studies have detected rotational lines of various molecules at mm wavelengths \citep[most notably CO, e.g.,][]{dutrey96,dutrey03}, but the brightest of these lines are invariably optically thick and therefore probe only the surface layers of the disc.  Velocity maps of molecular lines in the brightest sources appear consistent with Keplerian rotation \citep[e.g.,][]{simon00}, but these too are currently limited to relatively small samples.

By contrast, cold dust ($\sim 10$K) in the outer disc ($\gtrsim 50$AU) is readily observed in (sub-)millimetre continuum emission \citep[e.g.,][]{beckwith90}.  This emission is thought to be optically thin, so by assuming a dust opacity we are able to convert an observed continuum flux into a total dust mass.  Standard practice then assumes a canonical dust-to-gas ratio (typically 1:100) in order to infer a total disc mass.  Disc masses measured in this manner range from around a Jupiter mass up to $\sim 0.1$M$_{\odot}$ \citep{aw05}, and while these observations are subject to significant systematic uncertainties \citep[notably in the assumed opacity and dust-to-gas ratio,][]{hartmann06} they provide a valuable means of surveying large numbers of TT discs \citep[e.g.,][]{beckwith90,aw05,ec06}.  In addition, interferometric observations of mm continuum emission have resolved many discs in nearby star-forming regions, and the typical sizes of isolated discs are found to be $\sim100$AU \citep[e.g.,][]{aw07}.

\subsection{Observations of disc evolution}\label{sec:evo_obs}
Having discussed the various means by which protoplanetary discs are observed, I now discuss how we use such observations to constrain models of disc evolution.  We cannot observe the evolution of individual objects, as the evolutionary timescales are much too long, so we instead draw information on evolution from statistical studies of disc populations.  Timescales are generally derived through the use of stellar ages, obtained by fitting observations of TTs to pre-main-sequence stellar evolutionary models \citep[e.g.,][]{dm94}.  Stellar ages derived in this manner are subject to significant systematic uncertainties \citep[e.g.,][]{tlb99,baraffe02}, but the relative ages of different stellar populations are usually robust.

As mentioned above, the most commonly-used tool in the study of circumstellar discs is near-infrared photometry.  Large scale surveys of young stellar clusters provide a good base for statistical studies, and these studies remain one of the richest sources of data for disc study.  A summary of the work of many such surveys is presented in the oft-cited work of \citet{haisch01}, which presents the fraction of sources with excesses (and therefore discs) as a function of cluster age.  At stellar ages of less than 1Myr more than 90\% of objects are found to show infrared excesses, and the data are consistent with all such stars being born with circumstellar discs.  However by ages of around 10Myr very few significant infrared excesses are detected, and thus the typical lifetimes of inner dust discs are constrained to be a few Myr \citep{haisch01,sic06a}.  Moreover, (sub-)mm continuum observations of cold dust in the outer disc correlate very well with observations of inner dust discs \citep{aw05}, which suggests that the disc lifetimes inferred from observations of the inner disc are indeed representative of the entire disc.  Observations of gas are much less common, but measurements of accretion rates for individual objects show a steady decline in accretion rate with stellar age, with the decline occurring over timescales of a few Myr \citep{hcga98,muz00,calvet_ppiv}.  

\begin{figure}[t]
	\begin{center}
        \resizebox{\hsize}{!}{
        \includegraphics[angle=270]{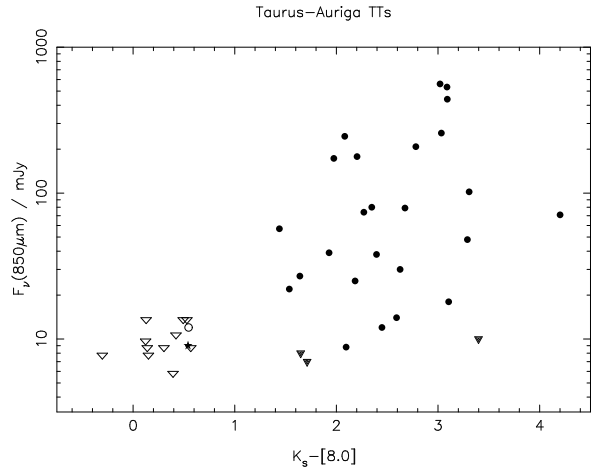}
        }
	\end{center}
        \caption{IR and sub-mm observations of TTs in Taurus-Auriga.  The horizontal axis shows the infrared colour, in magnitudes, measured between the 2MASS $K_{\mathrm s}$ (2.2$\mu$m) and {\it Spitzer} IRAC 8.0$\mu$m photometric bands.  The vertical axis shows the SCUBA 850$\mu$m flux, which is thought to correlate well with disc mass: circles represent detections, while triangles denote 3-$\sigma$ upper limits.  As in Fig.~\ref{fig:hartmann05}, filled symbols represent CTTs, open symbols WTTs, and the star denotes CoKu Tau/4.  All of the accreting CTTs show strong IR excess, and the only CTTs not detected by SCUBA are known to be discs viewed edge-on.  By contrast all of the WTTs show photospheric colours, and only two are detected in the SCUBA sample: a known transition object (CoKu Tau/4) and a known binary.  This suggests that the rapid inner disc clearing inferred from IR observations applies equally to the cold outer disc.  [Figure adapted from \citet{acp06b}, using data from \citet{hartmann05} and \citet{aw05}.]}
        \label{fig:scuba_ir}
\end{figure}

However these observations allow us to constrain more than just the lifetimes.  In surveys of low-mass clusters (i.e., clusters with no massive stars) it has long been known that very few objects are found {\it between} the Class II and Class III states \citep{skrutskie90,kh95,hartmann05}.  If we look at the infrared colour-colour diagrams of such clusters there is a striking gap between the Class II and Class III loci, with few (if any) objects observed in the intermediate region (see Fig.~\ref{fig:hartmann05}).  Observations in the mid-infrared show a similar gap \citep{persi00,bontemps01,padgett06}, and this suggests that the transition between the CTT and WTT states is extremely rapid.  This conclusion is drawn from the fact that very few objects are ``caught in the act'' of clearing their discs, and the observations constrain the dispersal time to be 1--2 orders of magnitude shorter than the lifetime \citep{sp95,ww96}.  This behaviour is clearly not consistent with the steady decline predicted by accretion disc theory \citep[][also the chapter by Lodato in this volume]{lbp74,hcga98}; instead the inner dust observations show an extremely rapid ``shut-off'' in the disc emission.  Morover, observations of the outer disc show a similar lack of objects between the CTT and WTT states \citep{duvert00,aw05}, suggesting that the entire disc is dispersed almost simultaneously.  This is summarized in Fig.~\ref{fig:scuba_ir}, which shows that accretion signatures, infrared excess and sub-mm continuum emission all disappear contemperaneously.  Therefore we require that the entire disc, from 0.1--100AU, is dispersed on a timescale of $10^4$--$10^5$yr after a lifetime of a few Myr.  Conventional accretion disc theory cannot explain such a rapid decline \citep{act99}, and so another explanation is required.

\subsection{Summary of disc observations}\label{sec:obs_sum}
Many more observations of discs have been made than those discussed here, but the observations discussed here are those most relevant to disc evolution.  These observations place notable constraints on models of disc evolution, which are summarized below:
\begin{itemize}
\item Disc lifetimes are typically a few Myr, with a large scatter.
\item CTTs and WTTs co-exist at the same age in the same clusters.
\item CTT disc masses range from $\sim0.1$M$_{\odot}$ to $\lesssim 0.001$M$_{\odot}$.
\item CTT accretion rates range from $\sim 10^{-7}$M$_{\odot}$yr$^{-1}$ to $\lesssim10^{-10}$M$_{\odot}$yr$^{-1}$.
\item Termination of (measurable) disc accretion is contemperaneous with disc clearing.
\item Disc clearing occurs rapidly (in $\sim10^5$yr), across the entire radial extent of the disc.
\end{itemize}
We see that all disc properties show a large scatter, and the origin of this scatter is not fully understood.  It is thought that disc evolution will result in a significant spread in observed properties \citep[e.g.,][]{hcga98}, and evolutionary models should be able to reproduce the range of parameters discussed here.  However, this cannot be the whole story, as we observe significant variation in disc parameters around stars of the same age in the same clusters \citep[e.g.,][]{muz00,sic06}.  This suggests that the initial conditions for disc evolution are not uniform, and we note in passing that realistic models should also attempt to take account of the disc formation process \citep[e.g.,][]{dnt06}.

As seen above, we are able to infer much about protoplanetary disc evolution from observations, but these observations are subject to significant caveats.  Most important is that fact that most of our knowledge of disc evolution comes from observations of dust in discs, but this dust represents only a small fraction of the total disc mass.  This dichotomy arises because the absorption and emission of radiation in the disc is overwhelmingly dominated by the trace dust component \citep[e.g.,][]{semenov03}: while gas dominates the disc mass, dust dominates the opacity.  Consequently we must take care when comparing models to observations, as many disc models treat only the gaseous component of the disc \citep[e.g.,][]{hcga98,cc01}.  In the following sections I will discuss gas and dust evolution in turn, attempting always to anchor theoretical discussion to observations.

\section{Evolution of gas discs}\label{sec:gas}
\subsection{Processes affecting gas disc evolution}\label{sec:gas_proc}
A number of dynamical processes affect the evolution of gas in protoplanetary discs, notably ``viscous'' evolution due to the transport of angular momentum \citep[e.g.,][see also the chapter by Lodato in this volume]{ss73,pringle81,bh98}, stellar magnetic processes such as disc winds and jets \citep[e.g.,][see also the chapter by Ferreira in this volume]{shu_ppiv,konigl_ppiv}, dynamical encounters with nearby stars \citep[e.g.,][]{cp93,sc01}, and photoevaporation of the disc, both by the central star and external illumination \citep[e.g.,][see discussion below]{holl94,jhb98}.  The relative importance of these processes was discussed in detail in the review of \citet{holl_ppiv}, which concluded that the dominant processes affecting disc evolution in the majority of systems are angular momentum transport (henceforth ``viscosity'') in the inner regions ($\lesssim 10$AU), and photoevaporation at larger radii.  The other processes discussed can dominate over short timescales or in small regions of the disc (such as magnetically-launched jets very close to the star), or can be important for a small fraction of stars (such as tidal stripping by nearby stars, which affects $\lesssim 10$\% of the stars in the Orion nebula cluster, \citealt{sc01}), but do not dominate the evolution of most TT discs.  I refer the reader to \citet{holl_ppiv} for a discussion of the various timescales associated with these different processes; here I focus on photoevaporation and viscous evolution.  I first outline the basics of photoevaporation theory, and then discuss models which combine photoevaporation and viscous evolution.  I conclude this Section by presenting a schematic model for the evolution of the gas in discs around (isolated) TTs, and discussing various aspects of current research in this area.

\subsection{Disc photoevaporation}\label{sec:photoevap}
The term ``photoevaporation'' refers to mass-loss from a circumstellar disc due to radiative heating of the disc material, either from the central star or by an external heat source.  The importance of photoevaporation was recognised as long ago as \citet{bs82}, but the first models which treated the flow and radiative transfer problems in a self-consistent manner were reported by \citet{holl93} and \citet{sjh93}, and then extended in detail by \citet{holl94}.  These models were applied with great success to ultracompact H{\sc ii} regions around massive stars, but in recent years it has become clear that photoevaporation is also important for low-mass stars, both in terms of disc evolution and planet formation.  Here I discuss the basic physics of photoevaporation, and its applications to the evolution of discs around TTs.

I first outline the basic principles of photoevaporation for the simplest case: that of a wind driven only by ionizing photons from the central star, in the absence of a significant stellar wind.  The reasons for choosing this example are twofold.  Firstly, it is the simplest case, and almost the only one with a straightforward analytic solution; secondly, it is (arguably) the case most applicable to T Tauri stars.  Most of what follows represents a summary of the work of \citet{holl94}, but for reasons of space and clarity I will adopt a somewhat simpler approach.  I refer the reader to \citet{holl94} for a more detailed analysis.

\begin{figure}[t]
	\begin{center}
        \resizebox{\hsize}{!}{
        \includegraphics[angle=270]{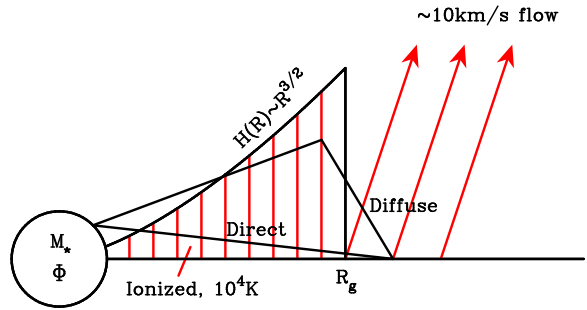}
        }
	\end{center}
        \caption{Schematic illustration of a photoevaporative disc wind.  Inside the gravitational radius $R_{\mathrm g}$ an ionized disc atmosphere is formed, with scale-height $H(R)$.  Outside $R_{\mathrm g}$ material is evaporated away from the disc.  (Figure adapted from \citealt{holl94,holl_ppiv}.)}
        \label{fig:holl00}
\end{figure}

Consider a star of mass $M_*$ with a surrounding circumstellar disc.  We assume that the disc is geometrically thin, and also that the disc is optically thick to Lyman continuum photons: both of these assumptions are valid for CTTs.  Thus, if the star produces a strong ionizing radiation field a thin ionized layer, similar to an H\,{\sc ii} region, is created on the disc surface.  The ionized gas has a temperature of $\simeq10^4$K, and therefore has much greater thermal energy (per particle) than the cold disc material beneath it.  Outside some radius the thermal energy of the ionized gas is larger than its gravitational potential energy, and the ionized gas is unbound.  As a result the ionized gas will flow away from the disc surface: this flow is known as a photoevaporative disc wind (see Fig.~\ref{fig:holl00}).

The characteristic length scale, known as the gravitational radius and denoted by $R_{\mathrm g}$, is found by equating the thermal and gravitational energies of the ionized gas (or alternatively by equating the local sound and orbital speeds), and is given by
\begin{equation}\label{eq:R_g}
R_{\mathrm g} = \frac{GM_*}{c_{\mathrm s}^2} = 8.9 \left(\frac{M_*}{\mathrm M_{\odot}}\right) \mathrm {AU}
\end{equation}
where $c_{\mathrm s}$ is the sound speed of the ionized gas, typically 10km s$^{-1}$.  The flow rate outside $R_{\mathrm g}$ depends on the density at the base of the ionized layer, and to first order the mass-loss rate per unit area is simply evaluated as $\rho c_{\mathrm s}$ at the base of the ionized region (where $\rho$ is density).  The density at the base of the ionized region depends on the radial and vertical structure of the disc atmosphere, and consequently we can reduce the evaluation of the wind rate to a radiative transfer problem.  In making this assumption we have neglected the flow structure of the wind, but as we will see below this makes only an order-of-unity difference to the calculations.  Moreover, as we are considering only ionizing radiation the (extremely complicated) radiative transfer problem is reduced to a calculation of ionization equilibrium.

Inside $R_{\mathrm g}$ the scale-height of the ionized atmosphere is much larger than the scale-height of the (cold) disc.  We can therefore approximate the structure of the ionized layer as that of an isothermal disc in hydrostatic equilibrium, with scale-height $H(R)$ given by
\begin{equation}
H(R) =  \left(\frac{c_{\mathrm s}}{v_{\mathrm K}}\right) R = \left(\frac{c_{\mathrm s}^2 R^3}{GM_*}\right)^{1/2} \quad , \quad R < R_{\mathrm g}
\end{equation}
where $v_{\mathrm K}$ is the Keplerian orbital velocity.  The sound speed (temperature) of the ionized gas is constant, so we can substitute from Equation \ref{eq:R_g} to find a simpler form for the scale-height:
\begin{equation}
H(R)=R_{\mathrm g}\left(\frac{R}{R_{\mathrm g}}\right)^{3/2} \quad , \quad R < R_{\mathrm g}.
\end{equation}
As mentioned above, the critical factor in determining the flow rate is the (number) density at the base of the ionized region, $n_0(R)$.  \citet{holl94} showed that the attenuation of stellar Lyman continuum photons through the disc atmosphere is very high, and that the diffuse (recombination) field dominates for all radii of interest\footnote{Note that when hydrogen undergoes radiative recombination approximately 1/3 of recombinations are to the ground state, and therefore emit another ionizing photon.}.  If we assume that the diffuse flux incident on the disc surface at a radius $R$ is dominated by the disc atmosphere very close to $R$ (an assumption verified by the numerical work of \citealt{holl94}), then we can adopt a Str\"omgren condition, equating ionizations with recombinations at the base of the ionized layer:
\begin{equation}
\alpha_{\mathrm B} R^3 n_0^2(R) \propto \Phi \quad , \quad R < R_{\mathrm g} \, .
\end{equation}
In this equation $\alpha_{\mathrm B} = 2.6\times10^{-13}$cm$^3$s$^{-1}$ is the Case B recombination coefficient for atomic hydrogen at $10^4$K \citep{allen}, $\Phi$ is the stellar ionizing luminosity (in units of photon s$^{-1}$), and the proportionality accounts for the difference between the diffuse radiation field at $R$ and the stellar ionizing flux $\Phi$.  Thus
\begin{equation}\label{eq:static}
n_0(R) = n_{\mathrm g} \left(\frac{R}{R_{\mathrm g}}\right)^{-3/2} \quad , \quad R < R_{\mathrm g} \, ,
\end{equation} 
where $n_{\mathrm g}$ is the base density at $R_{\mathrm g}$.  We calculate $n_{\mathrm g}$ by means of a Str\"omgren criterion similar to that above, so
\begin{equation}\label{eq:n_g}
n_{\mathrm g} = C \left(\frac{3 \Phi}{4\pi \alpha_{\mathrm B} R_{\mathrm g}^3}\right)^{1/2}
\end{equation}
where the order of unity constant $C$ reflects the difference in intensity between the stellar and diffuse ionizing fields.  The numerical work of  \citet{holl94} fixes the value as $C \simeq 0.14$.

At radii beyond $R_{\mathrm g}$, in the ``flow'' region, the ionized layer flows at approximately the sound speed.  There is no longer a direct line-of-sight from the star to the top of the disc atmosphere, so the irradiation beyond $R_{\mathrm g}$ is dominated by the diffuse field produced at $R_{\mathrm g}$.  Consequently, at radii beyond $R_{\mathrm g}$ the incident flux is simply the number of recombinations at $R_{\mathrm g}$, corrected for geometric dilution.  If we equate ionizations and recombinations at the base of the disc atmosphere for $R>R_{\mathrm g}$, we find
\begin{equation}
\alpha_{\mathrm B} R_{\mathrm g}^3 n_{\mathrm g}^2 \left(\frac{R_{\mathrm g}}{R}\right)^2 =  \alpha_{\mathrm B} R^3 n_0^2(R) \quad , \quad R>R_{\mathrm g}
\end{equation}
and therefore
\begin{equation}\label{eq:flow}
n_0(R) = n_{\mathrm g}\left(\frac{R}{R_{\mathrm g}}\right)^{-5/2} \quad , \quad R > R_{\mathrm g}.
\end{equation}
The mass-loss per unit area from the disc, $\dot{\Sigma}_{\mathrm {wind}}(R)$, is given by the product of the mass density and the sound speed, so
\begin{equation}\label{eq:sig_prof}
\dot{\Sigma}_{\mathrm {wind}}(R) = 2 \rho_0(R) c_{\mathrm s} = 2 n_0(R) c_{\mathrm s} \mu m_{\mathrm H}\quad , \quad R > R_{\mathrm g}
\end{equation}
where $m_{\mathrm H}$ is the mass of one hydrogen atom, $\mu$ is the mean molecular weight of the gas (taken to be $\mu=1.35$) and the factor of 2 accounts for mass-loss from both sides of the disc.  There is no mass-loss within $R_{\mathrm g}$ (i.e.,~$\dot{\Sigma}_{\mathrm {wind}} = 0$ for $R < R_{\mathrm g}$.), and the mass-loss profile from the disc falls off as $R^{-5/2}$ outside $R_{\mathrm g}$.  The integrated mass-loss rate is found by computing
\begin{equation}
\dot{M}_{\mathrm {wind}} = \int_{R_{\mathrm g}}^{\infty} 2\pi R \dot{\Sigma}_{\mathrm {wind}}(R) dR \, ,
\end{equation}
and if we substitute the form for $\dot{\Sigma}_{\mathrm {wind}}(R)$, integrate, and re-write in terms of parameters typical of TTs, we find that
\begin{eqnarray}\label{eq:wind_rate}
\dot{M}_{\mathrm {wind}} \simeq 4.4\times 10^{-10} \left(\frac{\Phi}{10^{41}\mathrm s^{-1}}\right)^{1/2} \\
\nonumber
\times \left(\frac{M_*}{1\mathrm M_{\odot}}\right)^{1/2} \mathrm M_{\odot} \mathrm {yr}^{-1}.
\end{eqnarray}
Thus the mass-loss profile is entirely specified by just 2 parameters: the stellar mass and ionizing flux.  Moreover, the rapid fall-off in the mass-loss rate per unit area at radii beyond $R_{\mathrm g}$ means that mass-loss very close to $R_{\mathrm g}$ dominates over mass-loss at larger radii.  

This simplified scheme captures the qualitative processes well, but more recent work has resulted in a number of quantitative improvements to the model.  The presence of dust in the ionized gas can alter the radiative transfer problem \citep[although this is more applicable to massive stars than to TTs, e.g.,][]{ry97}, and the effects of non-ionizing UV radiation can also be important \citep[e.g.,][see also Section \ref{sec:proplyds} below]{jhb98}.  Detailed consideration of the hydrodynamic properties of the wind have also resulted in improvements to the wind models.  The isothermal wind can be considered as a Bernoulli flow \citep[see, e.g., the discussion in ][]{dull_ppv}, and when pressure gradients are taken into we find that the ``critical radius'' at which the wind is launched is in fact $\simeq R_{\mathrm g}/5$ \citep{liffman03}.  Additionally, the hydrodynamic models of \citet{font04} showed that the transition between the ``static'' and ``flow'' regions is not as sharp as the step-function assumed by \citep{holl94}, and find that the wind rate is reduced by a factor of $\simeq 3$ (primarily due to the somewhat smaller, sub-sonic, launch velocity).  As mentioned above, the qualitative behaviour of the wind is unchanged, but the reader should note that I will make use of the wind models of \citet{font04} in subsequent calculations.

\subsubsection{External irradiation: the ONC proplyds}\label{sec:proplyds}
As mentioned above, early photoevaporation models tended to focus on massive stars \citep[e.g.,][]{holl94,ry97}, and only in more recent years has the application of these models to TTs become popular.  However, it has long been known that photoevaporation of TT discs due to {\it external} irradiation can be important, and this is especially relevant to the case of the so-called ``proplyds'' in Orion\footnote{The term ``proplyd'' was originally coined as an abbreviation of PROtoPLanetarY Disc, following observations of such discs in silhouette against the background emission in the Orion nebula \citep{odell93}.  However, more recently this terminology has become disfavoured in the literature, and the term ``proplyd'' is now generally used to refer only to discs of this specific type: discs that are being photoevaporated by the radiation from nearby massive stars.}.  These are a small number ($\sim100$) of TTs with discs observed in silhouette (i.e.,~in absorption) against the diffuse background emission in the Orion nebula, and are typically surrounded by characteristic cometary-shaped nebulae \citep{odell93,mo96}.  They are also bright in emission lines such as H$\alpha$ and [N{\sc ii}], and were quickly recognized as discs which are being photoevaporated by the UV emission from the nearby O-stars at the centre of the cluster.  \citet{jhb98} constructed detailed models of this process, and these models had great success in explaining the observed emission.  In these models the wind is primarily driven by non-ionizing, far-ultraviolet (FUV), photons, as the ionizing (EUV) photons are absorbed in the flow well above the disc surface.  The wind rates are large, $\sim10^{-7}$M$_{\odot}$yr$^{-1}$ (although this depends strongly on the proximity of the proplyd to the ionizing source), and the characteristic cometary shape is explained by the interaction between the photoevaporative flow and the stellar wind from the nearby O-star(s).  However, the combination of this large wind rate and the measured disc masses gave rise to the so-called ``proplyd lifetime problem'': in the presence of such winds, the typical disc lifetimes should be so short ($\sim 10^5$yr) that we should not observe many, if any, proplyds.

A neat solution to the proplyd lifetime problem has been proposed by \citet{adams04} and \citet{clarke07}.  \citet{adams04} noted that the gravitational radius for FUV-driven flows is typically greater than the outer disc radius (as FUV-heated winds have much lower temperatures, $\sim100$--1000K, than the EUV-heated winds discussed above).  They constructed hydrodynamic models of the flow, and found that the wind rate is in fact a strong function of the disc size.  If the disc is larger than $\simeq 100$AU in size then the outer radius is larger than the gravitational radius, and the wind rate is large.  However, smaller discs have little or no material outside the gravitational radius and consequently produce weaker winds (launched radially from the outer disc edge) for the same external irradiation.  \citet{clarke07} coupled the wind profiles of \citet{adams04} to a simple viscous disc evolution model, and found that discs tend to tend to evolve towards a quasi-steady state where the wind rate matches the viscous spreading of the disc.  In this model small discs rapidly expand until the wind becomes efficient enough to limit the disc size, while large discs are rapidly photoevaporated down to smaller sizes.  The dominant factor controlling disc evolution is the initial disc mass, and \citet{clarke07} find that a simple spread of initial disc masses can explain the observed distribution of disc sizes in Orion, solving the proplyd lifetime problem.  However, some uncertainties remain in this picture, notably that the theoretically-predicted mass-loss rates \citep[from, e.g.,][]{adams04} are still significantly smaller than those inferred from spectroscopic observations of proplyd flows \citep[e.g.,][]{hod99}.  One one hand it seems that the low mass-loss rates predicted by models solve the proplyd lifetime problem; on the other hand, that the higher rates derived from observations suggest that the proplyd lifetime problem still exists.  Future studies, both observational and theoretical, will hopefully resolve this discrepancy.  Note also that, as mentioned in Section \ref{sec:gas_proc}, external irradiation of this kind affects only a small fraction of protoplanetary discs, but that the consequences for those discs which are affected can be spectacular.

\subsection{Photoevaporation plus viscous evolution}
Recently, models which incorporate both viscous evolution and photoevaporation by the central star have become popular \citep[e.g.,][]{cc01,mjh03,ruden04,tcl05,acp06b}.  Models of this type show a number of attractive properties: here I review the basic principles of this class of models, first outlined by \citet{cc01}.

\subsubsection{Basic concepts}
\begin{figure}[t]
	\begin{center}
        \resizebox{\hsize}{!}{
        \includegraphics[angle=270]{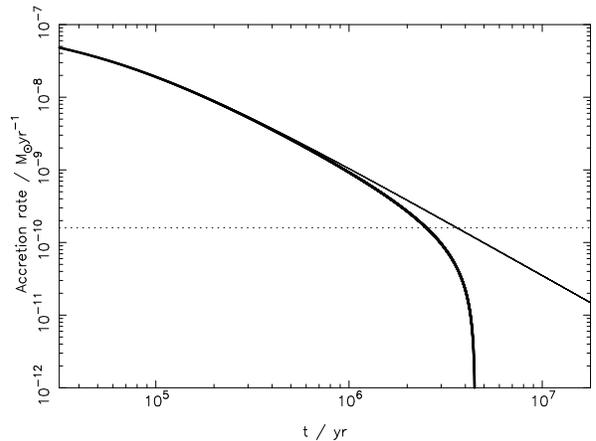}
        }
	\end{center}
        \caption{Decline of accretion rate (on to the star) in different disc models.  The thin solid line shows the evolution of the similarity solution of \citet{hcga98}, with the fiducial photoevaporative wind rate shown as a dotted line.  The heavy solid line shows the accretion rate when a photoevaporative wind is incorporated into the disc evolution model: the behaviour follows the similarity solution at early times, but the wind results in a rapid termination of accretion once the disc accretion rate falls to the level of the wind.  (Figure adapted from \citealt{cc01}, using the models of \citealt{acp06b}.)}
        \label{fig:cc01}
\end{figure}

Conventional models of disc evolution \citep[e.g.,][see also the chapter by Lodato in this volume]{lbp74,hcga98} give rise to a power-law decline in the disc surface density with time, and a similar decline in the mass accretion rate.  Thus while such models are broadly consistent with observations of TTs, they are unable to reproduce the rapid clearing observed at the end of the disc lifetime (see Section \ref{sec:evo_obs}).  The basic premise of the model proposed by \citet{cc01} is that, at some late stage in the evolution of a disc, the accretion rate through the disc falls to a level comparable to the rate of mass-loss due to photoevaporation.  The photoevaporative mass-loss is concentrated near the gravitational radius, so at this point in the evolution the outer disc, beyond $R_{\mathrm g}$, is unable to re-supply the disc inside $R_{\mathrm g}$ (as all of the accreting material is ``lost'' to the wind).  Once deprived of re-supply the inner disc de-couples from the outer disc, and viscosity drains the inner disc on to the star.  This occurs on the viscous timescale of the {\it inner} disc (which is much shorter than the disc lifetime, see Section \ref{sec:timescales} below), and the inner disc therefore disappears very rapidly, in a manner consistent with the many of the observations described in Section \ref{sec:evo_obs} (see Fig.~\ref{fig:cc01}).  To date this is the only class of models to have reproduced the observed two-timescale behaviour, and due to the role of the wind in precipitating this behaviour this model was christened the ``UV-switch''.

\subsubsection{The ``direct wind''}\label{sec:direct}
This class of models has a number of attractive properties, but the scenario presented by \citet{cc01} is not valid after the inner disc has drained.  The wind models described in Section \ref{sec:photoevap} assume that the disc is optically thick to ionizing photons at all radii, but this assumption is no longer valid once the inner disc has drained to a sufficiently low level.  Once the inner disc drains the ionized atmosphere disappears, eliminating the diffuse radiation field, and instead we must consider the consequences of direct irradiation of the inner disc edge.  This process was modelled in detail by \citet{acp06a}; here I summarize the basic results.

Much insight into this problem can be gained from an analysis similar to that in Section \ref{sec:photoevap}: neglecting the hydrodynamic properties of the wind, and reducing the problem to one of ionization/recombination balance.  We again assume that the mass-loss per unit area from the disc at a given point is given by $\rho c_{\mathrm s}$, and again seek to evaluate the density at the base of the ionized layer.

We neglect recombinations between the source and the disc surface, and also assume azimuthal symmetry (integrating over the azimuthal coordinate $\phi$ throughout).  Therefore, along any given line-of-sight from the source the rate of recombinations, $N_{\mathrm {rec}}$, in a volume $\Delta V$ at the ionization front must balance the rate of ionizing photons absorbed at the front, $N_{\mathrm {ion}}$.  A column with polar angle $\theta$ and angular size $\Delta\theta$ has an area of $2\pi r^2 \sin \theta \Delta \theta$, so for an ionizing flux $\Phi$ the ionization rate at the front is\footnote{To avoid confusion I adopt the notation $(r,\theta,\phi)$ for spherical coordinates and $(R,z,\phi)$ for cylindrical: upper-case $R$ represents cylindrical radius, while lower-case $r$ denotes spherical radius.}
\begin{equation}
N_{\mathrm {ion}} = \frac{1}{2}\sin \theta \Delta \theta \Phi \, .
\end{equation}
We make the simplifying assumption that the volume $\Delta V$ has a thickness equal to the disc scale-height $H$ \citep[as perturbations to the disc on scales shorter than $H$ are unlikely to be dynamically stable, e.g.,][]{lpf85} so $\Delta V$ is given by
\begin{equation}
\Delta V = 2\pi R H \frac{r \Delta \theta}{\sin \beta} \, ,
\end{equation}
where $\beta$ is the angle between the ray-path and the ionization front.  The cylindrical radius $R=r\sin\theta$, so the recombination rate is
\begin{equation}
N_{\mathrm {rec}} = 2\pi \alpha_{\mathrm B} n_0^2 R H \frac{R \Delta \theta}{\sin \theta \sin \beta} \, .
\end{equation}
We then equate the ionization and recombination rates and re-arrange to find
\begin{equation}\label{eq:base_den}
n_0(R) = \left(\frac{\Phi \sin^2 \theta \sin \beta}{4\pi \alpha_{\mathrm B} R^2 H(R)}\right)^{1/2} \, .
\end{equation}
We see, therefore, that in general $n_0(R)$ depends on the geometry of the ionization front, which in turn depends on the density structure of the unperturbed disc.  This problem requires numerical solution, but it is instructive to frame the problem in terms of dimensionless scaling constants.

\begin{figure}[t]
\centering
        \resizebox{\hsize}{!}{
        \includegraphics[angle=90]{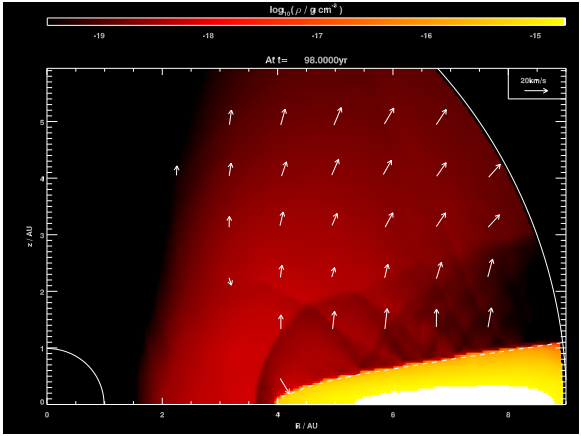}
        }
        \caption{Snapshot of one of the hydrodynamic wind simulations of \citet{acp06a}.  The gas density is plotted as a colour scale, with the grid boundaries denoted by solid lines and the ionization front by a dashed line.  Velocity vectors are plotted at regular intervals.}
        \label{fig:snapshot}
\end{figure}

We again evaluate the wind mass-loss rate per unit area as
\begin{equation}
\dot{\Sigma}_{\mathrm {direct}}(R,t) = 2 \mu m_{\mathrm H}n_0(R,t) v_{\mathrm {l}}(R,t)  \, ,
\end{equation}
The launch velocity $v_{\mathrm {l}}(R,t)$ is of order the sound speed of the ionized gas, and we have already made the simplifying assumption that at the inner disc edge $\Delta V$ has a radial thickness comparable to the vertical scale-height.  Along the disc midplane we have $\sin \theta = \sin \beta = 1$, so the inner edge density $n_{\mathrm {in}}$ can be expressed as
\begin{equation}\label{eq:n_in}
n_{\mathrm {in}} = C \left(\frac{\Phi}{4\pi \alpha_{\mathrm B} (H/R)_{\mathrm {in}} R_{\mathrm {in}}^3}\right)^{1/2} \, ,
\end{equation}
where $R_{\mathrm {in}}$ is the radius of the inner disc edge and $C$ is an order-of-unity scaling constant.  We then assume that $n_0(R)$ can be related to $n_{\mathrm {in}}$ by a dimensionless shape function $f\left(R/R_{\mathrm {in}}\right)$, so
\begin{equation}
n_0(R) = n_{\mathrm {in}} f\left(\frac{R}{R_{\mathrm {in}}}\right) \, .
\end{equation}
If we write then the launch velocity as
\begin{equation}\label{eq:launch_vel}
v_{\mathrm {l}}(R) = D c_{\mathrm s} \, ,
\end{equation}
where $D$ is another order-of-unity scaling constant, the mass-loss profile takes the form
\begin{equation}\label{eq:prof_form}
\dot{\Sigma}_{\mathrm {direct}}(R,t) = 2 CD \mu m_{\mathrm H} c_{\mathrm s}  n_{\mathrm {in}}(t) f\left(\frac{R}{R_{\mathrm {in}}(t)}\right)\, .
\end{equation}
The total mass-loss rate is found by integrating this from $R_{\mathrm {in}}$ to some outer disc radius $R_{\mathrm d}$, and if we substitute for $n_{\mathrm {in}}$ from Equation \ref{eq:n_in} and re-scale to parameters typical of TTs (taking $(H/R)_{\mathrm {in}}=0.05$) we find that
\begin{eqnarray}\label{eq:mout_anal}
\dot{M}_{\mathrm {direct}}(R_{\mathrm d}) = 2.35\times10^{-9} \, CD \, \left(\frac{\Phi}{10^{41}\mathrm s^{-1}}\right)^{1/2}  
\nonumber \\
\times  \left(\frac{R_{\mathrm {in}}}{3\mathrm{AU}}\right)^{1/2} \int_1^{R_{\mathrm d}/R_{\mathrm {in}}} x f(x) dx \, \mathrm M_{\odot}\mathrm{yr}^{-1} \, .
\end{eqnarray}
We have seen in Section \ref{sec:disc_obs} that TT discs typically have sizes $\gtrsim100$AU, so we expect that $R_{\mathrm d} \gg R_{\mathrm {in}}$.  Further, if we assume that the integral converges as $R_{\mathrm d}/ R_{\mathrm {in}} \rightarrow \infty$ (i.e.,~that the form of $f(x)$ falls off faster than $x^{-2}$), then we see that the mass-loss rate from the wind scales as $R_{\mathrm {in}}^{1/2}$.  Consequently, we expect the mass-loss rate to increase as the inner edge of the disc evolves outwards.  \citet{acp06a} conducted a series of hydrodynamical simulations of this problem (see Fig.~\ref{fig:snapshot}), and found that this analytic form does indeed provide a good fit to the wind profile.  The constants $CD$ and $a$ were found to depend weakly on the disc thickess $H(R)$: typical values (for $H/R=0.05$) are $CD=0.235$ and $a=2.42$.  Note also that the wind rate is around an order of magnitude larger than that of the diffuse-field wind, primarily due to the more efficient radiative transfer process.  Consequently, we expect the direct wind to dominate over the diffuse wind once a gap has opened in the inner disc.

\subsubsection{Evolutionary timescales}\label{sec:timescales}
Having established the behaviour of the photoevaporative flow during the different phases of disc evolution, we can now consider the evolutionary timescales that result.  [Similar analyses were presented by \citet{cc01} and \citet{acp06b}.]  Here we make use of the similarity solution presented by \citet{hcga98}, neglecting factors of order unity, and for simplicity adopt a linear viscosity law ($\nu \propto R$).  Consequently the local viscous timescale in the disc is given by
\begin{equation}
t_{\nu}(R) = \frac{R^2}{\nu(R)}
\end{equation}
and the characteristic scaling time for the evolution of the disc is 
\begin{equation}
 t_0 = \frac{R_0^2}{\nu_0} \, ,
 \end{equation}
 where $R_0$ is the characteristic radius of the similarity solution (typically $\simeq 10$AU for TTs, \citealt{hcga98,cc01}), and $\nu_0$ is the value of the viscosity at radius $R_0$.  When considering the evolution of the disc, there are three relevant timescales: 
\begin{enumerate}
 \item The ``lifetime'' of the disc (denoted by $t_{\mathrm {disc}}$): the time taken for the accretion rate to fall from its initial value to the level where photoevaporation becomes significant. 
 \item The ``inner clearing'' or ``gap-opening'' timescale ($t_{\mathrm {inner}}$): the time required for viscous draining of the inner ($\lesssim R_{\mathrm g}$) disc after the photoevaporative wind creates a gap in disc \citep{cc01}.
 \item The ``outer clearing'' timescale ($t_{\mathrm {outer}}$): the time required for the (direct) wind to clear the outer ($\gtrsim R_{\mathrm g}$) part of the disc \citep{acp06b}.
 \end{enumerate}
The disc lifetime can be estimated simply from the similarity solution, and is given by
\begin{equation}
t_{\mathrm {disc}} = t_0 \left(\frac{\dot{M}_{\mathrm {acc}}(0)}{\dot{M}_{\mathrm {wind}}}\right)^{2/3} \, ,
\end{equation}
where $\dot{M}_{\mathrm {acc}}(0)$ is the initial disc accretion rate.  The inner clearing timescale is simply the viscous timescale at $R_{\mathrm g}$, so
\begin{equation}
t_{\mathrm {inner}} = t_0 \left(\frac{R_{\mathrm g}}{R_0}\right) \, .
\end{equation}
If we neglect viscous evolution of the disc, the time required to clear the outer disc to a radius $R$ can be estimated as the disc mass inside $R$ divided by the direct photoevaporation rate.  The disc mass in turn is given by the product of the local viscous timescale and the accretion rate, so
\begin{equation}
t_{\mathrm {outer}} = \frac{M_{\mathrm d}(<R)}{\dot{M}_{\mathrm {direct}}(R)} = t_{\nu}(R) \frac{\dot{M}_{\mathrm {acc}}}{\dot{M}_{\mathrm {direct}}(R)}
\end{equation}

The accretion rate at which the gap opens is constant (approximately the diffuse wind rate), so the outer clearing timescale varies only with $\dot{M}_{\mathrm {direct}}(R)$.  We saw in Section \ref{sec:direct} that $\dot{M}_{\mathrm {direct}}(R)\propto R^{1/2}$, and by noting that the time required to clear the disc out to $R_{\mathrm g}$ is simply $t_{\mathrm {inner}}$ we see that
\begin{equation}
t_{\mathrm {outer}} = t_{\nu}(R) \left(\frac{R}{R_{\mathrm g}}\right)^{-1/2} \, .
\end{equation}
$(R/R_{\mathrm g})^{-1/2}<1$, so the direct wind clears the outer disc on a timescale shorter than the local viscous timescale (verifying our earlier decision to neglect viscosity)\footnote{Note that this remains true as long as $\dot{M}_{\mathrm {direct}}(R)$ is an increasing function of $R$.}.  We also see that the outer clearing time is longer than the inner clearing timescale by a factor of $(R/R_{\mathrm g})^{1/2}$: it takes somewhat longer to clear the outer disc than the inner disc.  In this scenario the ratio of the clearing timescale to the disc lifetime is therefore
\begin{eqnarray}
\frac{t_{\mathrm {outer}}}{t_{\mathrm {disc}}} = \frac{t_{\nu}(R_{\mathrm d})}{t_{\nu}(R_0)} \left(\frac{R_{\mathrm d}}{R_{\mathrm g}}\right)^{-1/2} \left(\frac{\dot{M}_{\mathrm {acc}}(0)}{\dot{M}_{\mathrm {wind}}}\right)^{-2/3} \\
\nonumber
= \frac{R_{\mathrm d}}{R_0} \left(\frac{R_{\mathrm d}}{R_{\mathrm g}}\right)^{-1/2} \left(\frac{\dot{M}_{\mathrm {acc}}(0)}{\dot{M}_{\mathrm {wind}}}\right)^{-2/3} \, .
\end{eqnarray}
If we select parameters typical of TTs we find that $t_{\mathrm {disc}}$ is typically a few Myr, and the ratio of the clearing timescale to the disc lifetime is 1--5\%.  Thus models which combine viscous evolution with photoevaporation of the disc evolve on timescales consistent with the observational constraints discussed in Section \ref{sec:obs_sum}.

\subsection{Schematic picture of gas disc evolution}\label{sec:gas_schematic}
\begin{figure}[t]
\centering
        \resizebox{\hsize}{!}{
        \includegraphics[angle=270]{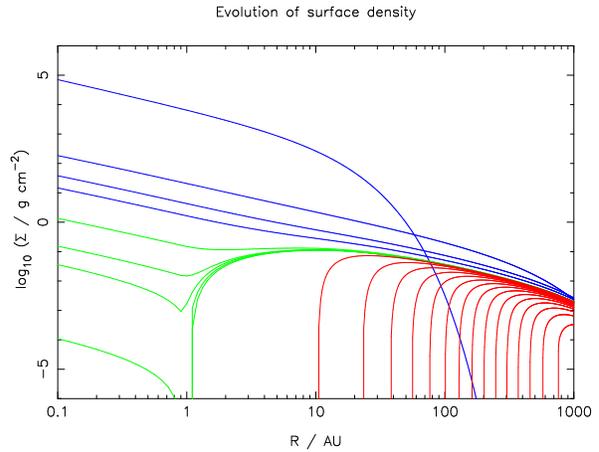}
        }
        \caption{Snapshots of disc surface density evolution from the fiducial model of \citet{acp06b}, colour-coded to identify the three stages of evolution described in the text.  Snapshots are plotted at $t=0$, 2, 4, 5.9Myr (blue), 6.0, 6.01, 6.02, 6.03Myr (green), 6.04, 6.05\ldots6.18Myr (red).}
        \label{fig:disc_model}
\end{figure}
We now have a model for the evolution of the gas disc over the entire disc lifetime, and we can describe it as ``three-stage'' model for gas disc evolution (see Fig.~\ref{fig:disc_model}):
\begin{enumerate}
\item {\bf Viscous phase: few Myr.} In this phase, which encompasses the majority of the disc lifetime, the photoevaporative wind is negligible and the disc evolves due to viscous transport of angular momentum.  Most of the disc mass is accreted on to the star, and most of the angular momentum is transported to large radii.
\item {\bf Gap-opening phase: $\mathbf {\lesssim10^5}$yr.} In this phase the (diffuse) photoevaporative wind cuts of the inner disc from re-supply, and the inner disc drains on its own, short, viscous timescale.
\item {\bf Clearing phase: few $\mathbf {10^5}$yr.} In this phase the inner disc is optically thin to ionizing radiation, and the disc is dispersed from the inside-out by the (direct) photoevaporative wind.
\end{enumerate}
This model has a number of attractive properties and, as discussed above, is the only model to date which satisfies the observed constraints on evolutionary timescales.  However, many uncertainties remain, and here I discuss some of the most important.

Perhaps the biggest uncertainty in models of disc photoevaporation is the UV luminosity of the central star.  The models discussed here require that the star produce a large flux of ionizing photons (the fiducial value of $\Phi=10^{41}$photon s$^{-1}$ corresponds to a luminosity of $\simeq 10^{30}$erg s$^{-1}$), and moreover require that the ionizing luminosity is sustained throughout the disc lifetime.  If the ionizing flux declines with time (as it would if it was powered by accretion) then the photoevaporative wind rate also declines, and the resulting disc evolution is inconsistent with the observational constraints discussed in Section \ref{sec:obs_sum} \citep{mjh03,ruden04}.  However, if the ionizing luminosity is produced in the stellar chromosphere, and is therefore driven by stellar magnetic activity rather than disc accretion, it is expected to be approximately constant over the lifetime of the disc.  X-ray emission from TTs (which is magnetically-driven) is uncorrelated with disc properties, and in fact some of the most X-ray luminous TTs are disc-less WTTs \citep{coup}.  The solar chromosphere emits Lyman continuum photons at $\sim10^{38}$photon s$^{-1}$ \citep[e.g.,][]{basri79,ayres97} and, given that TTs are much more magnetically active than the Sun, it is not unreasonable to expect TTs to have significantly larger ionizing luminosities.  Unfortunately direct observations of TTs in the UV are extremely difficult to make, and consquently there are few direct estimates of $\Phi$ available in the literature.  \citet{ks04} used a scaling argument to estimate the UV-spectrum of a young, active G-type star, and their spectrum has an ionizing flux of $2.5\times10^{41}$photon s$^{-1}$.  \citet{acp05} used archival UV observations to reconstruct chromospheric emission measures and estimate $\Phi$ for a small number of CTTs, and found $\Phi \sim 10^{42}$photon s$^{-1}$ to be typical.  They also found that the He{\sc ii} 1640\AA~line, thought to be produced by radiative recombination, was uncorrelated with any disc properties, suggesting that the ionizing emission is essentially unchanged during the TT phase.  These observations, however, are limited to a very small sample size, and given the lack of observational facilities currently available in the UV it is unclear when better such observations will be possible.

However, a great deal of indirect evidence has also emerged to support the hypothesis that TTs produce a strong ionizing flux.  \citet{font04} modelled the emission-line properties of the photoevaporative flow, and found that their model successfullly reproduced observed emission-line properties of several CTTs.  More recently, observations in the infrared have identified a number of emission lines that are thought to arise from radiative recombination in the disc atmosphere, most notably the [Ne{\sc ii}] line at $12.81\mu$m.  This line has recently been detected in the mid-IR spectra of a number of TTs \citep{pasc07,lahuis07}, and can be indicative of ionization of disc material by either EUV  (Hollenbach \& Gorti, {\it in prep.}) or X-ray \citep{gni07} photons.  If the emission is due to EUV ionization, then the inferred ionizing fluxes are once again $\sim 10^{41}$photon s$^{-1}$ \citep{pasc07}; however, if the emission is X-ray driven then it is unlikely to be important for disc dynamics, as X-ray heating does not drive a significant disc wind \citep{acp04b}\footnote{Note, however, that while the EUV fluxes are uncertain, the X-ray emission from these objects is readily observed.  Consequently, any constraints on the EUV emission derived from [Ne{\sc ii}] emission are technically only upper limits, as some fraction of the line flux is due to the (known) X-ray emission.}.  \citet{esp07} found evidence for a correlation between [Ne{\sc ii}] emission and stellar accretion rate but no correlation between X-ray emission and [Ne{\sc ii}] emission, and interpreted this as evidence for (E)UV-heating of the disc atmosphere.  However, previous theoretical models have rejected accretion-shock emission as a mechanism for driving disc photoevaporation \citep{mjh03,acp04a}.  Also, the correlation covers only a small region of the observed parameter space, with a small sample size (7 detections), so its significance remains unclear.  Moreover, observations of disc emission lines in the mid-IR are hindered by the bright continuum emission from the disc, and consequently most of the detections of [Ne{\sc ii}] emission to date have been observed in so-called ``transitional'' discs, which are characterised by unusually weak continuum emission in the near- and mid-IR (see Section \ref{sec:trans}).  It seems likely, therefore, that these observations are subject to significant selection biases, and that larger samples, with better signal-to-noise, will be required to break the degeneracy between X-ray and EUV excitation.  A possible solution to this problem has recently been suggested by \citet{herczeg07}, who used high-resolution (echelle) spectroscopy to resolve the [Ne{\sc ii}] line in the nearby transitional disc TW Hya.  They measured an intrinsic line-width of $\simeq20$km s$^{-1}$, which strongly suggests that the line is formed in a hot disc atmosphere at a radius of a few AU.  (The TW Hya disc is observed very close to face-on, so the line-width cannot be explained by Keplerian rotation unless the emission arises very close to the star.)  The profile of the [Ne{\sc ii}] line will differ depending on whether it arises in a bound disc atmosphere (X-ray excitation) or a photoevaporative wind (EUV).  Unfortunately the observation of this single, face-on source does not break the X-ray/EUV degeneracy, but \citet{herczeg07} suggest similar observations of a sample of discs at different inclination angles should resolve this uncertainty.

Our discussion so far has largely neglected the influence of non-ionizing, FUV heating of the disc, primarily for reasons of clarity.  As we have seen in Section \ref{sec:proplyds}, however, FUV heating is capable of driving a significant wind from the outer parts of the disc, and should be considered in models of disc evolution.  The much more complex heating and cooling problem in this case means that simple models in general do not suffice, as the dominant heating and cooling mechanisms change with the local disc conditions.  The geometrically simpler case of FUV heating by external irradiation has been modelled in depth by \citet{jhb98} and \citet{adams04}, and was discussed in Section \ref{sec:proplyds}.  However, to date no self-consistent models of FUV photoevaporation by the central star have been published.  Work in progress was discussed by \citet{gh04} and \citet{dull_ppv}, and although the details remain uncertain the basic principles seem secure.  FUV irradiation tends to heat the disc to much lower temperatures than EUV heating (typically 100--1000K), so FUV driven winds tend to be launched from correspondingly larger radii ($\gtrsim100$AU).  Moreover, the FUV heating rate from the central star is likely dominated by the accretion-shock emission \citep{mjh03,acp04a}, so time-dependent models will behave differently from those of EUV photoevaporation.  It seems likely that FUV photoevaporation will have a strong influence on the sizes of discs around TTs, and may in principle remove a significant fraction of the disc mass.  However, FUV heating is unlikely to affect the disc significantly at radii $\lesssim 50$AU, and therefore will not affect the EUV-driven clearing processes discussed above.  In the context of our picture of disc evolution it seems likely that FUV-photoevaporation will act throughout the evolution, and may shorten the duration of the viscous phase of evolution by removing much of the disc mass.  However, these details remain uncertain, and will hopefully be addressed in the near future.

\subsection{Summary of gas disc evolution}\label{sec:gas_sum}
In this Section we have discussed the various processes that affect the evolution of gas discs, and considered in detail models of disc photoevaporation.  We have discussed external photoevaporation in massive clusters, and the application of these models to the proplyds observed in the ONC.  We have shown that models which combine photoevaporation by the central star with viscous evolution of the disc predict rapid disc clearing after a long lifetime, and shown that such models satisfy the observational constraints discussed in Section \ref{sec:obs}.  Lastly, we have discussed current and future research in this area, from both theoretical and observational perspectives.

\section{Dust dynamics and evolution}\label{sec:dust}
We now turn our attention to the solid component of the disc, generally referred to as ``dust''\footnote{In an astrophysical context the term ``dust'' refers to solid material across a very large size range, from sub-$\mu$m sized grains up to rocks of metre-size and larger.}.  Dust is thought to make up only around 1\% of the total mass of protoplanetary discs, but it is crucial to the study of disc evolution for several reasons.  Firstly, the magnetohydrodynamic (MHD) turbulence that is thought to drive angular momentum transport in accretion discs requires that the disc be ionized above some low level \citep{gammie96,bh98}.  The dominant source of ionization in the midplane of protoplanetary discs is thought to be ionization of heavy elements by high-energy photons and cosmic rays \citep{glassgold_ppiv,fromang02}, and in this region most of the heavy elements are ``locked up'' in dust grains.  In addition, free electrons in the disc are readily absorbed by dust grains, and consequently an understanding of the distribution of dust in the disc is crucial to our understanding of gas disc evolution.  Secondly, as discussed in Section \ref{sec:obs}, while dust represents only a small fraction of the mass in protoplanetary discs, it contributes most of the disc opacity.  As a result most observations of protoplanetary discs observe the dust rather than the gas, so ideally models which seek to reproduce such observations should treat the dust and gas components of the disc independently.  Additionally, most TT discs are primarily heated by stellar irradiation, and as the dust dominates the disc opacity it is therefore crucial to our understanding of the thermal structure of discs.  Lastly, and most importantly for this chapter, the dust in discs is thought to represent the ``building blocks'' of planets.  Any theory of planet formation must explain the growth of dust from the small, sub-$\mu$m grains found in the interstellar medium to the very large solid bodies and planetary cores we see in planetary systems.  The evolution and dynamics of dust in discs is therefore of crucial importance to our understanding of both disc physics and planet formation.

\subsection{Basic physics of grain growth}\label{sec:dust_growth}
We first address the basic physics of dust growth, using a very simple approach.  Much of the discussion that follows is originally attributed to Victor Safronov, whose 1969 monograph {\it Evolution of the protoplanetary cloud and the formation of the Earth and the planets} still forms the basis of most modern theories of planet formation.  Here I discuss some of the simple arguments, first presented by Safronov, on the growth of grains in protoplanetary discs.

We can estimate the rate at which grains grow very simply, by considering collisions between dust grains in the disc.  We assume a typical material density of dust particles of $\rho_{\mathrm d}=1$g cm$^{-3}$, and note that typical midplane gas densities in TT discs are $\rho_{\mathrm g}\sim10^{-9}$g cm$^{-3}$.  If the grains are sufficiently small that they are very well coupled to the gas by aerodynamic drag (see Section \ref{sec:dust_motion} below), then the velocity dispersion, $\sigma$, of the grains is approximately that of the Brownian motion of the particles, which can be approximated as
\begin{equation}
\sigma \simeq \sqrt{\frac{m_{\mathrm H}}{m}} c_{\mathrm s} \, .
\end{equation}
If we assume that small particles stick with 100\% efficiency when they collide, the mean growth rate of particle mass is simply given by
\begin{equation}
\frac{dm}{dt} = \pi s^2 \rho_{\mathrm g} Z \sigma \, .
\end{equation}
Here $s$ is the radius of the dust particles, which are assumed to be spherical, and $Z$ is the dust-to-gas ratio (metallicity) in the disc.  The mass of a grain $m = (4/3)\pi s^3 \rho_{\mathrm d}$, so the rate of growth of particle size can be written as
\begin{equation}
\frac{ds}{dt} = 4 \frac{\rho_{\mathrm g}}{\rho_{\mathrm d}} Z \sigma \, .
\end{equation}
The typical sound speed in a disc at 1AU is $c_{\mathrm s} \simeq 1$km s$^{-1}$ (which gives $\sigma \simeq 0.1$cm s$^{-1}$), and if we assume a canonical dust-to-gas ratio of $Z=0.01$ we find a growth rate of
\begin{equation}
\frac{ds}{dt} \simeq 10^{-4} \mathrm {cm}\, \mathrm {yr}^{-1} \, .
\end{equation}
Given that the grains move relative to the gas (see below), the velocity dispersion $\sigma$ is likely to be significantly larger than we have estimated above.  (Strictly, this estimate applies only to the smallest grains, with $s\lesssim0.1$$\mu$m.)  Consequently, even if we assume that grains do not stick with 100\% efficiency we still expect to see rapid growth of small grains, with grains growing to cm-size on a timescale $\lesssim10^4$yr.  However, once grains reach this size range they no longer remain well-coupled to the gas in the disc, and this analysis breaks down.  Instead we must consider the effects of gas-drag forces on the motions of the dust grains: forces which produce some surprising results.

\subsection{Motion of dust grains}\label{sec:dust_motion}
We now consider the motion of dust particles in protoplanetary discs.  Such grains are subject to gravity, centrifugal forces and aerodynamic drag, and so we will now discuss the effects of aerodynamic drag on dust grains in discs.  A spherical particle of radius $s$, moving at a velocity $v$ relative to gas of density $\rho_{\mathrm g}$ experiences an aerodynamic drag force which opposes its motion:
\begin{equation}
F_{\mathrm D} = -\frac{1}{2}C_{\mathrm D} \pi s^2 \rho_{\mathrm g} v^2 \, .
\end{equation}
This expression has three terms: the cross-sectional area of the grain, $\pi s^2$; the ram pressure exerted on the grain, $\rho_{\mathrm g} v^2$; and the drag coefficient $C_{\mathrm D}$.  The drag coefficient depends on the velocity of the grain relative to the gas, and also on the size of the grain relative to the mean-free-path of gas molecules $\lambda$ \citep{whipple72,w77}.  For small particles ($s \lesssim 1$cm), where the size of the particles is less than $\lambda$ [formally where $s < (9/4)\lambda$] we are in the Epstein regime, and the drag coefficient is given by
\begin{equation}
C_{\mathrm D} = \frac{8}{3}\frac{v_{\mathrm {th}}}{v} \, ,
\end{equation}
where $v_{\mathrm {th}} = (8/\pi)^{1/2} c_{\mathrm s}$ is the mean thermal velocity of the gas.  For larger particles we instead find ourselves in the Stokes' regime, where the drag coefficient depends on the Reynolds number of the flow.  In this regime $C_{\mathrm D}$ is typically approximated by a piecewise function of the Reynolds number, such as that given in \citet{w77}.

It is useful to define the stopping timescale (or drag timescale), $t_{\mathrm s}$, which is the timescale on which frictional drag will cause an order-of-unity change in the momentum of the dust grain (relative to the gas).  Here we define the stopping time as\footnote{Note that different authors tend to define the stopping timescale slightly differently, often with different numerical pre-factors.  Consequently care must be taken when comparing different works, as the exact definitions of $t_{\mathrm s}$ may change.}
\begin{equation}
t_{\mathrm s} = \frac{mv}{|F_{\mathrm D}|} \, ,
\end{equation}
so in the Epstein regime the stopping time is therefore given by
\begin{equation}\label{eq:t_stop}
t_{\mathrm s} = \frac{\rho_{\mathrm d}}{\rho_{\mathrm g}}\frac{s}{v_{\mathrm {th}}} \, .
\end{equation}
If we adopt values typical of conditions in a protoplanetary disc (e.g.,~$\rho_{\mathrm g}=10^{-9}$g cm$^{-3}$, $v_{\mathrm {th}}=1$km s$^{-1}$) and the standard dust density of $\rho_{\mathrm d}=1$g cm$^{-3}$, we see that the stopping time for $\mu$m-sized grains is of order seconds.  This is much smaller than the typical dynamical (orbital) timescale, so we see that small grains are extremely well-coupled to the gas.  However, metre-size particles have $t_{\mathrm s} \sim \Omega^{-1}$ and are only marginally coupled to the gas, and much larger bodies are mostly unaffected by gas drag.

\subsubsection{Settling}
The simplest motion of dust grains we can consider is that of vertical settling in an isothermal gas disc.  We estimate the settling timescale by equating the opposing forces of drag and gravity.  For vertical displacements $z \ll R$, the vertical component of the gravitational force on a grain is simply
\begin{equation}
F_{\mathrm g} = -m \Omega_{\mathrm K}^2 z \, ,
\end{equation}
where $\Omega_{\mathrm K} = (GM_*/R^3)^{1/2}$ is the Keplerian orbital frequency.  If we equate this with the drag force (in the Epstein regime) and re-arrange we find a typical settling velocity of
\begin{equation}
v_{\mathrm {settle}} = \frac{\Omega_{\mathrm K}^2}{v_{\mathrm {th}}} \frac{\rho_{\mathrm d}}{\rho_{\mathrm g}(z)} s z\, ,
\end{equation}
we can then estimate the settling timescale as
\begin{equation}
t_{\mathrm {settle}} = \frac{z}{v_{\mathrm {settle}}} = \frac{v_{\mathrm {th}}}{\Omega_{\mathrm K}^2} \frac{\rho_{\mathrm g}(z)}{\rho_{\mathrm d}} \frac{1}{s} \, .
\end{equation}
For $\mu$m-sized particles in a typical protoplanetary disc the settling timescale is $\sim10^5$yr, but larger grains are expected to sediment towards the midplane in much shorter timescales.  Note also the dependence on the gas density: in a vertically isothermal disc $\rho_{\mathrm g} \propto \exp(-z^2/2H^2)$, so we expect particles to sediment out of the upper layers of the disc rapidly, but settle more slowly at smaller $z$.

In principle vertical settling will increase the midplane dust-to-gas ratio, and can result in increased dust collision (and growth) rates.  However, this simple analysis considers only a static, laminar vertical disc structure, and real TT discs are thought to be turbulent.  Settling of dust in a turbulent disc is a much more complex problem, as the turbulence can lift the grains to high $z$ on the eddy timescale of the turbulence, which is much shorter than the expected dust settling timescale.  Several recent studies have used numerical simulations to study the effects of MHD turbulence on dust settling \citep[e.g.,][]{jk05,csp05,turner06}, and while some details remain contested, all the models agree that turbulence does result in much slower settling than predicted above.  Indeed, it seems likely that the vertical distribution of small grains in discs is sustained by some sort of equilibrium between turbulent diffusion and sedimentation.  (For further discussion of the effects of turbulence see the chapter by Klahr in this volume.)

\subsubsection{Radial drift}\label{sec:rad_drift}
The equations governing the radial drift of dust in protoplanetary disc were first derived by \citet{w77}; here we follow the more recent analysis of \citet{tl02,tl05}.  We first derive the relationship between the rotational velocity of the gas, $v_{\phi,\mathrm g}$ and the Keplerian orbital velocity $v_{\mathrm K} = (GM_*/R)^{1/2}$.  We consider the radial component of the equation of motion thus:
\begin{equation}\label{eq:gas_speed}
\frac{v_{\phi,\mathrm g}^2}{R} = \frac{GM_*}{R^2} + \frac{1}{\rho_{\mathrm g}}\frac{dP}{dR} \, .
\end{equation}
Here the left-hand side is the centrifugal acceleration, which balances the accelerations due to gravity and gas pressure on the right-hard side.  If we assume that the gas pressure at the disc midplane can be approximated as a power-law
\begin{equation}
P \propto R^{-n}
\end{equation}
and assume a locally isothermal equation of state
\begin{equation}
P = \rho c_{\mathrm s}^2 \, ,
\end{equation}
then the pressure gradient term in Equation \ref{eq:gas_speed} becomes
\begin{equation}
\frac{1}{\rho_{\mathrm g}}\frac{dP}{dR} = -n \frac{c_{\mathrm s}^2}{R} \, .
\end{equation}
If we substitute this into Equation \ref{eq:gas_speed} and multiply by $R$, we find that
\begin{equation}
v_{\phi,\mathrm g}^2 = v_{\mathrm K}^2 (1-\eta) \, ,
\end{equation}
where the term
\begin{equation}
\eta = n \frac{c_{\mathrm s}^2}{v_{\mathrm K}^2}
\end{equation}
denotes how sub-Keplerian the gas is.  Typical power-law discs have values of $n\simeq 2.75$--3, depending on the choice of viscosity and temperature laws.  Therefore, if we assume that $H/R = c_{\mathrm s}/v_{\mathrm K} = 0.05$ and $n=11/4$ (consistent with an $\alpha$--disc with $T\propto R^{-1/2}$), we see that the gas in a protoplanetary disc is typically sub-Keplerian by 
\begin{equation}
(1-\eta)^{1/2} = 0.0034 \, .
\end{equation}
This may seem like a small fraction, but bear in mind that the Kepler speed for a 1M$_{\odot}$ star at 1AU is $\simeq30$km s$^{-1}$.  Consequently, we expect particles orbiting at the Kepler velocity to be subject to a very strong headwind, $\sim 100$m s$^{-1}$.

We now consider the radial drift of a dust particle of radius $s$ orbiting in such a gas disc.  Before we do, however, it is useful to define a dimensionless stopping time $T_{\mathrm s}$.  This is simply the stopping timescale $t_{\mathrm s}$ divided by the orbital timescale, and here we define $T_{\mathrm s}$ as\footnote{Note again that exact definitions of $T_{\mathrm s}$ differ in the literature, commonly by factors of $2\pi$.}
\begin{equation}
T_{\mathrm s} = t_{\mathrm s}\Omega_{\mathrm K} = t_{\mathrm s} \frac{v_{\mathrm K}}{R} \, .
\end{equation}
If we now consider the radial and azimuthal equations of motion (EoMs) for a dust grain, we see that
\begin{equation}\label{eq:rad_eom}
\frac{d v_{\mathrm {r,d}}}{dt} = \frac{v_{\phi,\mathrm d}^2}{r} - \Omega_{\mathrm K}^2 r - \frac{1}{t_{\mathrm s}} (v_{\mathrm {r,d}} - v_{\mathrm {r,g}})
\end{equation}
\begin{equation}\label{eq:az_eom}
\frac{d}{dt}(r v_{\phi,\mathrm d}) = -\frac{r}{t_{\mathrm s}}(v_{\phi,\mathrm d} - v_{\phi,\mathrm g} )
\end{equation}
Here the subscripts $_r$ and $_{\phi}$ denote the radial and azimuthal components of velocity respectively, with the additional subscripts $_{\mathrm g}$ and $_{\mathrm d}$ used to distinguish the gas and dust velocities.  In the radial EoM, the first term is the centrifugal acceleration, the second term the acceleration due to gravity, and the third term the frictional drag force (which opposes the motion of the dust grain, and is zero if the dust and gas move at the same velocity).  In the azimuthal direction the only acceleration is that due to the drag force.  We then make the simplifying assumption that the dust particles spiral in on approximately circular orbits (i.e.,~the specific angular momentum of a particle is always close to Keplerian), so that to first order
\begin{equation}
v_{\phi,\mathrm d} \simeq v_{\phi,\mathrm g} \simeq v_{\mathrm K} \, .
\end{equation}
Consequently we can simplify the azimuthal EoM by noting that
\begin{equation}
\frac{d}{dt}(r v_{\phi,\mathrm d}) \simeq v_{\phi,\mathrm d} \frac{d}{dr} (r v_{\mathrm K}) = \frac{1}{2} v_{\phi,\mathrm d} v_{\mathrm K} \, ,
\end{equation}
and by substituting into Equation \ref{eq:az_eom} and re-arranging we find that
\begin{equation}\label{eq:veldg}
v_{\phi,\mathrm d} - v_{\phi,\mathrm g}  \simeq -\frac{1}{2} \frac{t_{\mathrm s} v_{\mathrm K}}{r} v_{\mathrm {r,d}} = -\frac{1}{2} T_{\mathrm s} v_{\mathrm {r,d}} \, .
\end{equation}
We then make the second simplifying assumption that, to $O((H/R)^2)$, there is no net acceleration in the radial direction (i.e.~$d v_{\mathrm {r,d}}/dt \simeq0$).  If we re-write the Keplerian velocity in terms of the gas orbital velocity and $\eta$ thus
\begin{equation}
\Omega_{\mathrm K}^2 R = \frac{v_{\phi,\mathrm g}^2}{R} + \frac{\eta v_{\mathrm K}^2}{R} \, ,
\end{equation}
we can substitute for the appropriate term in the radial EoM to find
\begin{equation}
\frac{v_{\phi,\mathrm d}^2}{r} - \frac{v_{\phi,\mathrm g}^2}{r} - \eta \frac{v_{\mathrm K}^2}{r} - \frac{1}{t_{\mathrm s}} (v_{\mathrm {r,d}} - v_{\mathrm {r,g}}) = 0 \, .
\end{equation}
We expand the first two terms as a difference of squares (noting that $v_{\phi,\mathrm d} + v_{\phi,\mathrm g} \simeq 2v_{\mathrm K}$), and substitute the difference between the azimuthal velocities of the dust and gas from Equation \ref{eq:veldg}.  If we then multiply by $v_{\mathrm K}/R$ and re-write in terms of $T_{\mathrm s}$, we find that
\begin{equation}
v_{\mathrm {r,d}} T_s + v_{\mathrm {r,d}} T_s^{-1} = v_{\mathrm {r,g}} T_s^{-1} - \eta v_{\mathrm K} \, ,
\end{equation}
and therefore the drift velocity of the dust is given by
\begin{equation}\label{eq:driftvel}
v_{\mathrm {r,d}} = \frac{v_{\mathrm {r,g}} T_s^{-1} - \eta v_{\mathrm K} }{T_s + T_s^{-1}} \, .
\end{equation}
\begin{figure}[t]
\centering
        \resizebox{\hsize}{!}{
        \includegraphics[angle=270]{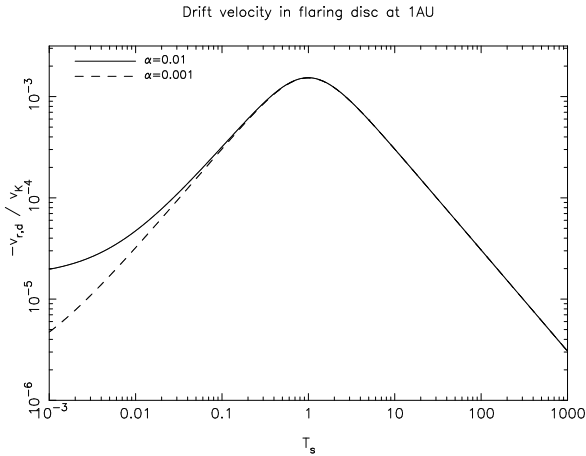}
        }
        \caption{Inward drift velocity, in units of the Kepler speed, for a typical TT disc at 1AU (see Equation \ref{eq:driftvel}).  The departure from symmetry is due to the inward gas velocity, which is computed for different $\alpha$ parameters at an accretion rate of $10^{-8}$M$_{\odot}$yr$^{-1}$.  Note the very large drift velocity for grains with $T_{\mathrm s} \sim 1$: for a 1M$_{\odot}$ star, $10^{-3}v_{\mathrm K} \simeq 3000$cm s${-1}$.}
        \label{fig:drift_vel}
\end{figure}
The form of this expression is shown in Fig.~\ref{fig:drift_vel}.  The radial velocity of the gas (due to viscosity) is typically very small, and if we neglect this motion we see that Equation \ref{eq:driftvel} is symmetric in $T_{\mathrm s}$.  The limit $T_{\mathrm s} \ll 1$ corresponds to very small particles ($s \lesssim 1$$\mu$m), and in this limit the grains are so well-coupled to the gas that they simply migrate inwards at the same speed as the gas.  By contrast, in the limit $T_{\mathrm s}\gg1$, which corresponds to very large particles ($s \gtrsim1$km), the particles have so much inertia that they are hardly affected by gas drag at all, and simply orbit at the Kepler speed\footnote{Note, however, that for $T_{\mathrm s} \gtrsim10$ we are in the (non-linear) Stokes regime, rather than the Epstein regime, so the simple linear relationship between particle size and stopping time (Equation \ref{eq:t_stop}) no longer holds.}.  The dynamics of bodies in this size regime, commonly referred to as ``planetesimals'', are in fact dominated by gravitational interactions between the planetesimals, and it is thought that gravitational focusing increases the collision rates between such bodies sufficiently that they can grow to planetary sizes very rapidly \citep[e.g.,][]{safronov69}.  

However, we also see that from Fig.~\ref{fig:drift_vel} that at the peak of the curve the inward drift is very rapid, reaching $\gtrsim 1000$cm s$^{-1}$ in a disc around 1M$_{\odot}$ star at 1AU.  At such velocities dust particles will spiral in to the central star in $\sim 100$yr, and for typical TT disc parameters, $T_{\mathrm s} =1$ corresponds to sizes of $s \simeq 50$--100cm.  We see, therefore, that in order to form planetesimals, particles must grow through the range of sizes that correspond to $T_{\mathrm s} \sim 1$ on very short timescales, and the conclusion is that planetesimal formation must occur very rapidly is unavoidable.  This is commonly referred to as the ``metre-size barrier'', and much modern research on planet formation focuses on how to solve this problem\footnote{Cynically one could say that if we did not know of the existence of planets, then the metre-size barrier would be a very robust argument as to why planets cannot form via the collisional growth of smaller particles!}.

In recent years several models have begun to combine models of gas evolution (similar to those discussed in Section \ref{sec:gas}) with radial drift of dust, with a view to modelling dust evolution over Myr timescales.  \citet{tcl05} combined the two-fluid model of \citet{tl02,tl05} with the photoevaporation/viscous model of \citet{cc01} to study the dynamics of mm grains, and found that, if unreplenished, the mm grain population in TT discs is accreted on to the star before the gas disc is cleared.  Dust-poor discs of this type are not seen \citep[e.g.,][]{aw05} but mm- to cm-size particles are observed  \citep{testi03,wilner05,rodman06}, so \citet{tcl05} concluded that some additional effect, such as grain growth or a reduced radial drift veocity, must slow the decline in the mm grain population.  \citet{aa07} extended the study of \citet{tcl05} to incorporate more realistic photoevaporation models, and also considered a range of grain sizes.  However, without considering grain growth their model was restricted to timescales of $\lesssim10^5$yr, so they focused on the behaviour of the dust during the clearing phase of the evolution (see Section \ref{sec:trans}).  More recently,  \citet{garaud07} considered models which include grain growth and dynamics, and found that the long-term evolution of the dust population depends very sensitively on the sizes of the grains that dominate the mass reservoir at large radius \citep[a result also found by][]{brauer07}.  Models of this type are becoming increasingly sophisticated, and provide a promising avenue for future research.

\subsection{Dust growth and planetesimal formation}
We have seen in Section \ref{sec:dust_growth} above that the growth of small dust grains can, in principle, occur very rapidly.  More realistic models than those presented above take into account more realistic physics of collisions and sticking, and in recent years much progress has been made, through both theoretical calculations and laboratory experiments \citep[e.g.,][see also the review by \citealt{dominik_ppv}]{bs04,dd05}.  Sticking efficiencies have been found to depend on a large number of factors, such as the shapes of the colliding bodies, the porosity of the grains, or the velocity of the collision.  In general it is found that sticking efficiencies tend to be high (10--100\%) at small grain sizes, and indeed the growth of small grains is so rapid that many models of dust growth require some form of growth-destruction equilibrium in order to sustain the observed population of $\mu$m-sized grains over Myr timescales \citep[e.g.,][]{dd05}.  However, sticking becomes much less efficient once grains reach the cm size range, resulting in much slower growth at these sizes.  Moreover, in this size range grain collisions can become destructive at speeds $\gtrsim 10$m s$^{-1}$, so the dynamics of the gas-grain interaction become important for growth processes also \citep{dominik_ppv}.

Many popular mechanisms for the formation of planetesimals are rooted in the so-called Goldreich-Ward instability \citep{gw73}.  In this scenario the vertical settling of small dust grains steadily increases the dust-to-gas ratio at the midplane, until eventually the thin dust sub-disc becomes gravitationally unstable and fragments.  We can estimate the conditions at which this occurs by a simple application of the \citet{toomre64} criterion, setting
\begin{equation}
Q = \frac{\sigma \Omega}{\pi G \Sigma_{\mathrm {dust}}} = 1 \, ,
\end{equation}
where $\sigma$ is the velocity dispersion in the dust layer and $\Sigma_{\mathrm {dust}}$ is the surface density of the dust layer.  If we assume that $\Sigma_{\mathrm {dust}} = \Sigma_{\mathrm {gas}}/100 \simeq 10$g cm$^{-2}$ and consider a 1M$_{\odot}$ star, we see that a velocity dispersion of $\sigma \lesssim 10$cm s$^{-1}$ is required in order for gravitational instability to fragment the dust layer.  Given that the gas sound speed is typically $\sim 1$km s$^{-1}$, we see that the dust sub-layer must be very thin indeed in order to become unstable.  However, this suggestion is attractive because it allows for the formation of planetesimals directly from small ($s\lesssim 1$mm) particles, and thus bypasses the size ranges what are most susceptible to radial drift.  

Unfortunately we can easily show that, in its simplest form, the Goldreich-Ward mechanism does not result in planetesimal formation.  We have seen above that the instability requires that the dust layer be approximately $10^4$ times thinner than the gas disc, and consequently at the disc midplane the local dust density dramatically exceeds the local gas density (by a factor $\sim 100$).  In this layer we therefore expect the solid material to dominate the dynamics, so gas pressure forces are negligible and the dust and gas both orbit at the Kepler speed.  Above this layer, however, the gas is significantly sub-Keplerian due to gas pressure (as discussed in Section \ref{sec:rad_drift} above), and there is therefore a large velocity shear in the vertical direction.  This shear is Kelvin-Helmholz unstable, and the resulting turbulence prevents $\sigma$ from becoming small enough for instability to set in \citep{cuzzi93}.  More recently, several sets of authors have argued that planetesimal formation can occur in this manner, if the local dust-to-gas ratio is significantly enhanced from the canonical value of 1:100 \citep[e.g.,][]{gl04}.  \citet{ys02} suggested that radial drift can result in large local enhancements in the dust-to-gas ratio, leading to instability, while \citet{tb05} suggested that photoevaporation by external irradiation (which preferentially removes gas from the disc) can likewise lead to gravitational instability in the dust layer.  More recently, \citet[][see also \citealt{yg05,yj07}]{jy07} showed that radial and azimuthal drift of particles in a gas disc can lead to the so-called streaming instability, and that the resulting structure in the gas disc can significantly increase the concentration of solid bodies.  They find that particles which are marginally coupled to the gas are most prone to clumping in this manner, and also find that the net radial drift of these particles is significantly reduced.  This mechanism provides an attractive solution to the metre-size problem, and represents one of the most promising means of forming planetesimals \citep{johansen07}.  

In a similar vein, much recent research has invoked structure in the gas disc in order to accelerate planetesimal formation by increasing collision rates.  A variety of similar models have been proposed, but all make use of the same underlying physics.  We saw in Section \ref{sec:rad_drift} that gas pressure gradients in protoplanetary discs result in radial migration of dust grains, but our treatment of the problem assumed a simple disc where the uniform pressure gradient invariably makes the gas slightly sub-Keplerian.  This results in inward migration of dust grains, but in general (as can be seen from Equations \ref{eq:gas_speed} \& \ref{eq:driftvel}), the dust will in fact move radially towards any local pressure maximum \citep[e.g.,][]{whipple72}.  If the gas disc has some local structure, such as spiral density waves, the gas orbits at slightly super-Keplerian speeds at the inner edge of the pressure enhancement.  The dust in this region instead migrates outwards, towards local pressure maxima, which can result in significant enhancements in the local dust-to-gas ratio.  Moreover, the particles which are most susceptible to this behaviour are the same particles that suffer the most severe radial drift, so models of this type offer an attractive solution to the metre-size problem.  This behaviour has been noted by a number of authors, and several different scenarios have been proposed.  \citet{rice04,rice06a} demonstrated that the spiral density waves induced by gravitational instability in the gas disc result in strong local enhancements of the dust-to-gas ratio in metre-size particles, and suggest that the enhancement may be large enough to allow planetesimal formation by gravitational interactions between the solid bodies.  \citet{pm04,pm06} observed similar behaviour in the spiral waves produced by the presence of a planet in the disc, and suggested that the presence of one planet may act to accelerate the formation of subsequent planets.  In addition, \citet{rice06b} and \citet[][see also \citealt{garaud07}]{aa07} both suggest that local pressure maxima can provide a ``size sorting'' mechanism in discs with gaps or holes, as grains with $T_{\mathrm s}\simeq1$ tend to clump in local pressure maxima, while smaller particles remain coupled to the the gas flow and larger particles orbit independently of the gas.  All of these models suggest that particle collision rates can be significantly enhanced {\it if} the gas disc has sufficient structure, but whether or not this occurs in real systems remains unclear.  In real discs any, all, or some combinations of these various mechanisms may apply, but the details of planetesimal formation remain rather poorly understood.

\subsection{Summary of dust evolution}
In this Section we have discussed how dust grains grow from small, sub-$\mu$m sizes to cm-size and larger.  We have considered the effects of aerodynamic drag forces on dust particles, and derived expressions for both the vertical settling and radial drift that result.  Particles of 10--1000cm are especially susceptible to radial drift, and we have seen that this ``metre-size barrier'' provides a significant challenge for theories of planetesimal formation.  We have discussed various mechanisms for forming planetesimals, but have noted the problems associated with all of them.  How to form planetestimals remains the biggest challenge for modern research in planet formation, and many of the details of this problem still elude us.

\section{Transitional discs}\label{sec:trans}
In recent years an increasing literature has built up on so-called ``transitional'' discs.  This small but important class of objects have properties between those typical of CTTs and WTTs, and as such are thought to represent discs which are observed during the (short) clearing phase of their evolution.  These objects are crucial to our understanding of disc clearing but, as the clearing process is rapid (see Section \ref{sec:evo_obs}), the objects are rather rare.  To date around 20 such objects are known, but this sample is expanding rapidly\footnote{Around a dozen new papers on transitional discs appeared while this article was being written!}.  In general it is believed that these objects show characteristics indicative of some degree of inner disc clearing, and different authors have invoked a variety of different physical mechanisms to explain these ``inner holes''.  In this section I attempt to summarize this rapidly-evolving subject, and try to highlight some key areas for future research in both observational and theoretical fields.

\subsection{What is a ``transitional'' disc?}\label{sec:trans_def}
Despite appearances, this is not a trivial question.  As mentioned above, a rough definition would encompass all objects with properties between those typical of CTTs and WTTs but, given the broad range of observations used to study protoplanetary discs, it should come as little surprise that a robust definition has so far been difficult to establish.  In general, objects have been classified as ``transitional'' based on their SEDs, and a loose definition would be that such objects typically show a flux deficit in the near- and mid-infrared, when compared to ``ordinary'' CTTs.  The first such objects were found by \citet{strom89}, who coined the term ``discs in transition'' and suggested that the observed SEDs were indicative of some degree of inner disc clearing.  Around 20 such objects are now known, but the advent of {\it Spitzer} is leading to a rapid increase in this sample size.  However, there is little consistency in the definitions used to classify these objects, in terms of either nomenclature or observational characteristics.   Recently, objects with similar properties have variously been called ``transitional'' \citep[e.g.,][]{calvet05,najita07}, ``passive'' \citep{mccabe06}, ``anaemic'' \citep{lada06} or ``cold'' \citep{brown07}, and a similar variety of observational selection criteria exist in the literature.

\begin{figure}
\centering
        \resizebox{\hsize}{!}{
        \includegraphics[angle=270]{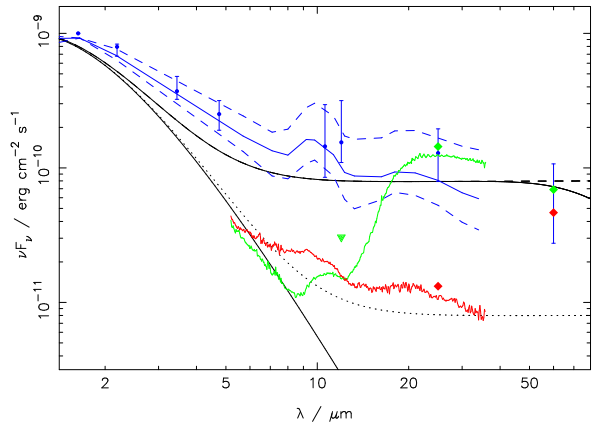}
        }
        \caption{Infrared spectra of transition discs.  The median SED of Taurus CTTs is shown in blue.  The photometric median SED from \citet{dalessio99} is shown as points with error bars, and the {\it Spitzer} IRS median SED from \citet{furlan06} is shown as solid and dashed lines: plotted in each case are the median and the quartiles of the observed distribution of SEDs.  The solid black lines show the stellar black body and simple disc model described in the text, and the dashed and dotted lines respectively show the optically thick and optically thin ($\tau=0.05$) limits of the disc model.  The SEDs of CoKu Tau/4 and FQ Tau are plotted in green and red respectively (data from \citealt{furlan06}, kindly provided in electronic form by Joan Najita): the lines are IRS spectra; the diamonds and triangles denote IRAS detections and upper limits respectively.}
        \label{fig:trans}
\end{figure}

The best-studied transition objects are the four nearby sources TW Hya \citep[e.g.,][found in the TW Hya accociation at a distance of $\simeq60$pc]{calvet02,wilner05}, GM Aur \citep[e.g.,][]{rice03,calvet05}, DM Tau \citep[e.g.,][]{calvet05} and CoKu Tau/4 \citep[e.g.,][all found in the Taurus-Auriga cloud at a distance of $\simeq140$pc]{forrest04,dalessio05}, and in essence these four objects have come to define the modern meaning of ``transitional disc''.  All four show little or no excess emission above their stellar photospheres in the near-infrared, but show excesses comparable to conventional CTTs at wavelengths longer than $\sim$10--20$\mu$m.  (A typical spectrum, that of CoKu Tau/4, is shown in Fig.~\ref{fig:trans}).  Detailed modelling has shown that the SEDs of these objects are best-fit by discs with large inner holes: that is, discs that are (mostly) cleared of dust at small radii, $\lesssim$5--10AU, but appear to be ``normal'' CTT discs at larger radii \citep{calvet02,dalessio05,calvet05}.  The best-fitting hole radii are typically a few to a few tens of AU, and do not appear to scale simply with any known stellar properties\footnote{Note, however, that the sizes of the holes are not especially well-known.  Recent mid-IR interferometric observations of TW Hya \citep{ratzka07} measured a significantly smaller inner hole radius than had previously been measured from the SED \citep{calvet02} or from mm interferometry \citep{hughes07}, and the origin of this discrepancy is not yet understood.}$^{\mathrm ,}$\footnote{Recently, \citet{cmc07} have claimed that the observed hole sizes for accreting transition discs correlate with the measured accretion rates, with $\dot{M} \propto \alpha r_{\mathrm {hole}}^2 / M_*$.  The small number of sources (three) and large error bars make the significance of this correlation difficult to assess, but  future observations should be able to either confirm or reject this hypothesis with high confidence.}.  A large (and increasing) number of different mechanisms have been proposed to explain these inner holes, with different studies variously invoking the presence of a planet \citep[e.g.,][]{rice03,quillen04}, photoevaporation \citep[e.g.,][]{acp06b}, dust evolution \citep[e.g.,][]{wilner05}, photophoresis \citep{krauss07} or enhancement of the magnetorotational instability \citep[MRI,][]{cmc07} to explain some or all of the observed properties.

However, when we look beyond the SEDs we see that observations of transitional discs show remarkable diversity.  For example, the observed (stellar) accretion rates and disc masses for these four objects alone differ by more than 3 orders of magnitude.  Moreover, the structures of these four discs seem to differ significantly within the holes.  GM Aur and DM Tau are both CTTs with significant gas accretion ($\sim10^{-8}$M$_{\odot}$yr$^{-1}$) observed at the stellar surface, but while GM Aur shows a weak near-infrared excess, suggestive of some optically thin dust close to the star, DM Tau appears to have a much ``cleaner'' hole in its dust disc \citep{calvet05}.  TW Hya is accreting at a much lower level ($\simeq 4\times10^{-10}$M$_{\odot}$yr$^{-1}$, \citealt{muz00}), but interferometric observations of the inner disc suggest the presence of a population of small dust grains that must be continually replenished, presumably by the destruction of larger bodies \citep{eisner06}.  CoKu Tau/4, by contrast, is a WTT with no detectable accretion, and shows no evidence for significant amounts of gas or dust within its hole.  

As discussed in Section \ref{sec:obs}, protoplanetary discs are generally extremely optically thick in the near- and mid-infrared, so the bulk of the disc emission at these wavelengths is expected to arise in a thin layer on the disc surface \citep[e.g.,][]{cg97}.  To first order, therefore, the infrared SED only tells us about the presence or absence of dust at a given temperature in the disc.  All of the models for disc clearing mentioned above result in similar infrared SEDs, and as such infrared observations alone are of limited use in distinguishing between models.  Indeed, no single object is unambiguously associated with any particular model.  Consequently it seems likely that a statistical approach, selecting objects on the basis of their SEDs and then investigating other disc properties, may be the most promising way of learning more about these objects.

The only such study published to date is that of \citet{najita07}, who used the {\it Spitzer} Infrared Spectrograph (IRS) survey of \citet{furlan06} to identify a sample of 12 transitional discs in the Taurus-Auriga cloud.  Taurus is uniquely suited to a statistical study of disc evolution, as the stellar properties of all of the sources are well known and the SCUBA survey of \citet{aw05} provides a uniform sample of disc masses for almost all of the known TTs.  \citet{najita07} utilised a demographic approach, and showed that their transitional objects show systematically lower accretion rates than CTTs of the same disc mass.  This strongly suggests that objects with transitional SEDs do represent an evolved state of CTTs, in agreement with previous studies of individual objects \citep[e.g.,][]{dalessio05,calvet05}.  However, the work of \citet{najita07} introduces a potential selection bias by selecting objects based on their infrared SEDs alone.  This is not necessarily problematic but, as the authors discuss, is not an ideal method of selecting a sample to compare with theoretical models.  Consequently, I will now attempt to outline a {\it theoretically} motivated definition of what a transitional disc is.

All of the theoretical models mentioned above (with the exception of dust evolution) produce deep holes in the inner (dust) disc, and a sharp transition between the inner, cleared region (the hole) and the outer disc.  Consequently I propose that the selection criterion used to define a transitional disc {\it for the purposes of comparison with theoretical models} should be one that unambiguously requires a hole in the dust disc.  Consequently objects should be classified as ``transitional'' based on the following criterion:
\\
\\
{\it Transition objects have discs which show optically thin emission at shorter wavelengths AND optically thick emission at longer wavelengths.}
\\
\\
Objects which satisfy this criterion must have radial disc structures which are clearly inconsistent with those typical of CTTs, and essentially must show strong evidence for a hole or gap in the inner regions of the disc.  In fact, for reasons of clarity, it may well make sense to class such objects as ``inner hole'' discs, essentially defining a sub-set of the more loosely identified transitional discs.  

An example of the application of this criterion is shown in Fig.~\ref{fig:trans}.  Shown in the figure is the median SED for CTTs in Taurus-Auriga \citep{dalessio99,furlan06}, as well as the SEDs of two transition disc taken from the sample of \citet{najita07}: CoKu Tau/4 and FQ Tau\footnote{These particular objects are chosen simply for illustrative purposes, as they have similar spectral types (M1.5 and M2 respectively) and luminosities.  FQ Tau is also known to be a binary system, but this does not affect this illustrative discussion.}.  Also shown is the spectrum of a stellar black-body (with a temperature of 3200K, typical of an early M star), and that of a simple disc model.  The disc model assumes power-law surface density, $\Sigma(R)$, and temperature, $T(R)$, profiles as follows:
\begin{equation}
\Sigma(R) = 100 \left( \frac{R}{1\mathrm {AU}}\right)^{-1} \mathrm g \, \mathrm {cm}^{-2} \quad , \quad R \le 100 \mathrm {AU}\, ,
\end{equation}
\begin{equation}
T(R) = 300 \left( \frac{R}{1\mathrm {AU}}\right)^{-1/2} \mathrm K \quad , \quad R \le 100 \mathrm {AU}\, .
\end{equation}
The minimum disc temperature is taken to be 10K, and the inner edge of the disc (i.e.,~the dust sublimation radius) is fixed at the point where $T(R)=1500$K.  The outer disc radius of 100AU is adopted merely for convenience, and does not affect the emission in the infrared.  The disc SED is then computed as
\begin{equation}
F_{\nu} = \frac{\cos i}{4\pi d^2} \int 2\pi R B_{\nu}\{T(R)\} \left[1-\exp\left(-\frac{\tau_{\nu}}{\cos i}\right)\right] dR \, ,
\end{equation}
where $d$ is the Earth-star distance (taken to be 140pc), $B_{\nu}\{T\}$ is the Planck function and $i=60^{\mathrm {\circ}}$ is the inclination angle of the disc.  The optical depth $\tau_{\nu}$ is evaluated as
\begin{equation}
\tau_{\nu} = \kappa_{\nu}\Sigma(R) \, ,
\end{equation}
assuming a simple power-law form for the dust opacity \citep{beckwith90}
\begin{equation}
\kappa_{\nu} = 0.1\frac{\nu}{10^{12}\mathrm {Hz}} \quad  \mathrm {cm}^2 \, \mathrm g^{-1} \, ,
\end{equation}
and the final SED is then computed by adding the contribution from the disc to the stellar black-body.

Also shown in Fig.~\ref{fig:trans} are the optically thick and optically thin ($\tau=0.05$) limits of the disc model.  In order to satisfy the criterion above, objects must have SEDs comparable to or below the optically thin limit at short wavelengths, but consistent with the optically thick limit at longer wavelengths.  [In practice, one could use the (normalised) quartiles of the median SED and the 8--10$\mu$m stellar (photospheric) flux to represent the optically thick and thin limits respectively.]  We see from Fig.~\ref{fig:trans} that the spectrum of CoKu Tau/4 shows essentially no excess over the stellar photosphere at wavelengths shorter than 8$\mu$m, but is consistent with an optically thick disc at wavelengths longer than $\simeq17$$\mu$m.  By contrast, the IRS spectrum of FQ Tau shows a weak excess across the entire IRS band, and is consistent with an approximately constant optical depth of $\tau \sim 0.1$.  The IRS spectrum of FQ Tau, therefore, does not require a genuine hole to be present in the disc, but could instead be fit by a disc with lower-than-average optical depth (possibly due to dust settling and/or growth).  In this case it seems likely that the disc does emit in the near-infrared, but that the relatively weak disc emission at such wavelengths results in a negligible excess over the stellar photosphere.

However, the IRAS 60$\mu$m observation of FQ Tau {\it does} show a significant excess, an order of magnitude above the optically thin limit, and consistent with the simple disc model at this wavelength.  This highlights what is potentially an important selection effect: that the wavelength range covered by the {\it Spitzer} IRS is only sensitive to fairly small holes, up to a few tens of AU in size\footnote{Given the large IRAS beam size this large 60$\mu$m excess could be due to a companion, but again this is not important for this illustrative discussion.}.  Any objects with larger inner holes ($\gtrsim 50$AU in size, although the cut-off depends strongly on stellar spectral type) will be defined as Class III objects on the basis of their IRS spectra alone, as the disc will only emit at wavelengths longer than those observed with {\it Spitzer}.  This highlights the need for multi-wavelength observations when identifying transitional discs, as techniques which are only sensitive to inner holes in a particular size range may bias subsequent interpretation of the data.  Consequently, I propose that future statistical studies identify inner hole sources using data that covers the widest possible range of wavelengths, as only in this manner will unbiased samples be guaranteed.

\subsection{Distinguishing between models of disc clearing}\label{sec:distinguish}
As mentioned above, a large number of models have been proposed to explain some or all of the observed properties of various transitional discs, and here I attempt to distinguish between some of these models.   If we use the selection criterion discussed above to select objects with inner holes, then it seems likely that the properties of these discs should reflect the manner in which the holes were formed.  I will focus on two mechanisms in particular: gap-opening by a massive planet embedded in the disc \citep[e.g.,][]{rice03,quillen04}, and gap-opening by the combined actions of photoevaporation and viscosity (as discussed in Section \ref{sec:gas}).  I restrict myself to considering these two models first for reasons of length, but also because these are the only models which tell the ``complete story''.  Mechanisms such as photophoresis \citep{krauss07} or MRI enhancement by X-rays \citep{cmc07} may well explain some of the observed properties of discs with inner holes, but these models require the action of some other mechanism to create the inner hole in the first place. 

\subsubsection{Gap-opening by planets}
The planet-disc interaction is discussed in much greater detail in the chapter by Klahr (this volume), but here I quickly review the criteria for a massive planet to open a gap in a disc.  If a planet is sufficiently massive then the tidal torques it exerts on the disc can open a gap close to the planet \citep[e.g.,][]{tml96}, and models of discs with such gaps predict SEDs consistent with those of several observed transition objects \citep[e.g.,][]{rice03,quillen04}.  Two criteria must be satisfied for a planet to open a gap in a disc: firstly, the tidal (Lindbland) torques exerted by the planet on the disc must overcome the local viscous torques in the disc; and secondly, the Hill radius (i.e.,~the region where the gravity of the planet is greater than that of the star) must be comparable to the disc thickness.  In the case of protoplanetary discs these criteria are approximately equivalent, and in an $\alpha$-disc the approximate criterion for gap-opening is \citep[e.g.,][]{tml96}
\begin{equation}
q \gtrsim \left(\frac{c_{\mathrm s}}{R_{\mathrm p} \Omega_{\mathrm p}}\right)^2 \alpha^{1/2}\, .
\end{equation}
Here $q$ is the ratio of the planet mass to the stellar mass, and the subscript $_{\mathrm p}$ denotes properties of the planet.  For typical parameters we find that planets with masses $\gtrsim 0.5$M$_{\mathrm {Jup}}$ will open a gap in the disc.  \citep[For a more detailed study of the gap-opening criterion see][]{edgar07a}.  Moreover, the criterion for gap-opening is effectively independent of the disc surface density (and therefore the disc mass), so a sufficiently massive planet will be able to open a gap in a disc of almost any mass.

Once the planet has opened a gap in the disc the tidal and viscous torques will clear the inner part of the disc, creating an inner hole \citep{quillen04,varniere06}.  Over long timescales the planet will then undergo Type II migration, but if the planet is sufficiently massive this timescale can be comparable to the disc lifetime \citep[e.g.,][]{sc95}.  However, the planet forms somewhat of a ``leaky barrier'', so unless the planet is extremely massive ($\gtrsim 10$M$_{\mathrm {Jup}}$) we expect some accretion of material, both on to the planet and beyond \citep[e.g.,][]{lsa99}.  The details of the planetary accretion process in discs with gaps remain somewhat uncertain \citep[see, e.g., the discussion in][]{lda06}, but it seems likely that most planet-induced gaps will have some accretion across the planetary orbit, and therefore some accretion on to the central star.

\subsubsection{Planets or photoevaporation?}
\begin{figure}
\centering
        \resizebox{\hsize}{!}{
        \includegraphics[angle=270]{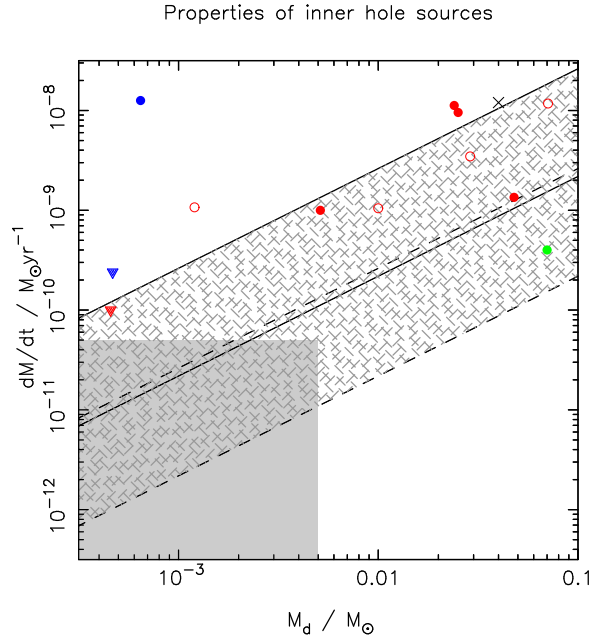}
        }
        \caption{Disc masses and accretion rates for discs with inner holes \citep[adapted from][]{aa07}.  The shaded region shows the allowed range of parameters for photoevaporative clearing, and the hatched region the range expected for a simple model of the planet-disc interaction.  The red and blue points show the transition discs identified by \citet{najita07}: red points are single stars, blue are binaries; circles represent detections, triangles upper limits.  The filled symbols show the objects which satisfy the selection criterion in Section \ref{sec:trans_def}; the open circles show those objects classed as transitional by \citet{najita07} which do not satisfy the more stringent selection criterion outlined here. Also shown are TW Hya \citep[green circle, data from][]{calvet05} and CS Cha \citep[black cross, data from][]{esp07}.  (Note also that the lower-left of the region allowed by the planet model seems rather unlikely on formation grounds, as in this region we require that the planet be more massive than the disc in which it is embedded.)}
        \label{fig:aa07_rev}
\end{figure}
We have seen above that planet-induced gaps tend to result in continued accretion interior to the planet, and can occur in discs with a wide range of masses.  By contrast, we saw in Section \ref{sec:gas} that photoevaporation opens a gap in the disc only once the accretion rate has dropped to a low level (that of the photoevaporative wind).  If we then assume some viscosity law in the disc this sets an upper limit on the masses of disc with ``photoevaporated'' holes \citep[$\simeq 5$M$_{\mathrm {Jup}}$, e.g.,][]{acp06b,aa07}.  Moreover, once viscosity has drained the inner disc we expect essentially no accretion on to the central star, as the direct wind clears the disc on a timescale shorter than the local viscous timescale (see Section \ref{sec:timescales}).  In light of these facts, \citet{aa07} and \citet{najita07} independently proposed that surveys of disc masses and accretion rates can be used to distinguish between these two mechanisms for inner disc clearing.

This is demonstrated in Fig.~\ref{fig:aa07_rev}, which combines the models of \citet{aa07} with the data of \citet{najita07}.  The hatched region on the plot shows the range of allowed parameters for a simple planet-disc model (adapted from \citealt{lsa99} and \citealt{lda06}), while the grey region marks the range expected for ``photoevaporated'' holes.  In the planet-disc model, the lines denote the loci where the parameters of the disc model are constant: the solid lines are for $\alpha=0.01$, while the dashed lines denote $\alpha=0.001$; in both cases the upper line shows the case of a 1M$_{\mathrm {Jup}}$ planet, while the lower line denotes a 12M$_{\mathrm {Jup}}$ planet.  (See discussion in \citealt{aa07} for more details.)  We see that several of the canonical transitional discs appear consistent with the embedded planet scenario, while only one or two of the observed sources fall in the range predicted by the photoevaporation scenario\footnote{Note, however, that there is considerable disagreement as to the significance of the upper limits on accretion rates in cases where no accretion is detected.  For example, \citet{sic06} assign upper limits of $10^{-12}$M$_{\odot}$yr$^{-1}$ to objects with no detectable UV excess and no broad H$\alpha$ lines, while the more conservative approach of \citet{najita07} assigns upper limits to such objects in the range $10^{-10}$--$10^{-9}$M$_{\odot}$yr$^{-1}$.  Robust upper limits to the accretion rates in objects categorised as WTTs will provide valuable additional constraints on models of such systems.}.  However, Fig.~\ref{fig:aa07_rev} also highlights the importance of selection biases in analyses of this type.  The theoretical arguments above consider the formation of deep holes or gaps in the inner disc, so it is essential that only objects which unambiguously possess such holes be considered in the analysis.  If we consider the 12 transition objects identified by \citet{najita07}, we see that 8--10 of them seem consistent with the planet-scenario (depending on how one treats the binaries in the sample), while only 2/12 are consistent with the photoevaporation scenario.  If we apply the selection criterion suggested in Section \ref{sec:trans_def}, however, we reject 4 of the single stars in the \citet{najita07} sample, and instead find that 4/6 objects appear to be planet-induced holes, while 2/6 are consistent with photoevaporation.  Moreover, we saw in Section \ref{sec:trans_def} that the wavelength range used to identify current samples of transitional discs may bias the samples towards smaller hole sizes ($\lesssim 50$AU).  Given that the timescale required for a planet to open a gap in a disc   is typically $\sim1000$ local orbital periods \citep[e.g.,][]{varniere06,edgar07b}, it seems unlikely that planets at radii beyond a few tens of AU will open significant gaps in their discs over the Myr lifetimes of TT discs\footnote{Additionally, models of planet formation and migration suggest that most gas-giant planets form at radii $\simeq5$--10AU \citep[e.g.,][]{armitage07b}.}.  By contrast, disc clearing by photoevaporation can produce inner holes of any size greater than $\simeq1$AU, and while the expected hole size distribution depends on the viscosity law adopted it is likely biased weakly towards larger hole sizes \citep{aa07}.  Given the small numbers of transition discs known, great care must be taken in applying selection criteria to the observed samples, and robust samples with uniform selection criteria are essential if we are to use such statistical methods to constrain mechanisms of disc clearing.

At present the data on transitional discs do not allow us to distinguish clearly between the different theoretical scenarios, but these data suggest that more than one mechanism may produce the inner holes we observe.  As discussed above, the survey of \citet{najita07} suggests a planet:photoevaporation hole ratio between 6:1 and 2:1 (based on very small number statistics).  The only comparable sample of accretion rates in transitional discs is that of \citet{sic06}, who identified 17 transitional discs in the Tr37 cluster.   Unfortunately Tr37 is too distant for TT disc masses to be measured, but \citet{sic06} found that roughly half their sample (9/17) are accreting, while half (8/17) showed no evidence for any on-going accretion.  As we observe more transitional discs these sample sizes will grow, and it seems such studies will provide vastly improved statistics within the next few years.

\subsection{Summary of transitional discs}
In this Section we have discussed transitional discs, which are thought to be discs observed between the CTT and WTT states.  We have described the observed properties of these objects by considering the best-known sources, and discussed the various theoretical mechanisms that have been proposed to explain their origin.  Infrared observations alone do not discriminate strongly between different models, and it seems likely that statistical studies of disc properties beyond the SEDs offer the most promising way of discriminating between these models.  Lastly, we have discussed various observational criteria used to identify these objects, including proposing a new, theoretically-motivated criterion, and highlighted the strong influence that selection effects can have on statistical surveys of transitional discs.


\subsection*{Acknowledgements}
I thank Phil Armitage and Cathie Clarke for critical readings of various drafts of this manuscript, and also thank them both for many stimulating discussions.  In addition, I thank Anders Johansen for a number of useful critical comments.  I am grateful that I was able to attend Scott Tremaine's recent lectures on planet formation at Leiden Observatory, as these helped greatly in organizing my thoughts when writing Section \ref{sec:dust}.  Lastly, I thank Joan Najita for providing me with the data used to make Figs.~\ref{fig:trans} \& \ref{fig:aa07_rev}, for a valuable discussion on transition discs, and for a thorough and insightful referee's report.

This work was supported by NASA under grants NAG5-13207 and NNG05GI92G from the Origins of Solar Systems and Beyond Einstein Foundation Science Programs, and by the NSF under grants AST--0307502 and AST--0407040.


\end{document}